\documentclass[aps,pre,twocolumn,superscriptaddress,longbibliography,hidelinks]{revtex4-2}

\usepackage{color}
\usepackage{graphicx}
\usepackage{siunitx}
\usepackage{amsmath}
\usepackage{hyperref}
\usepackage{bm}

\begin{document}

\title{Unified description of viscous, viscoelastic, or elastic thin active films on substrates}

\author{Henning Reinken}
\email{henning.reinken@ovgu.de}
\affiliation{Institut f\"ur Physik, Otto-von-Guericke-Universität Magdeburg, Universitätsplatz 2, 39106 Magdeburg, Germany} 

\author{Andreas M. Menzel}
\email{a.menzel@ovgu.de}
\affiliation{Institut f\"ur Physik, Otto-von-Guericke-Universität Magdeburg, Universitätsplatz 2, 39106 Magdeburg, Germany}

\date{\today}

\begin{abstract}
It is frequent for active or living entities to find themselves embedded in a surrounding medium.
Resulting composite systems are usually classified as either active fluids or active solids.
Yet, in reality, particularly in the biological context, a broad spectrum of viscoelasticity exists in between these two limits. There, both viscous and elastic properties are combined. 
To bridge the gap between active fluids and active solids, we here systematically derive a unified continuum-theoretical framework. It covers viscous, viscoelastic, and elastic active materials. %
Our continuum equations are obtained by coarse-graining a discrete, agent-based microscopic dynamic description. 
In our subsequent analysis, we mainly focus on thin active films on supporting substrates. Strength of activity and degree of elasticity are used as control parameters that control the overall behavior. 
We concentrate on the analysis of transitions between spatially uniform analytical solutions of collective migration. 
These include isotropic and polar, orientationally ordered states. A stationary polar solution of persistent directed collective motion is observed for rather fluid-like systems. It corresponds to the ubiquitous swarming state observed in various kinds of dry and wet active matter. 
With increasing elasticity, persistent motion in one direction is prevented by elastic anchoring and restoring forces. As a consequence, rotations of the spatially uniform migration direction and associated flow occur.  
Our unified description allows to continuously tune the material behavior from viscous, via viscoelastic, to elastic active behavior by variation of a single parameter. Therefore, it allows in the future to investigate the time evolution of complex systems and biomaterials such as biofilms within one framework. 
\end{abstract}

\maketitle

\section{Introduction}

Active matter is intrinsically out of equilibrium due to the continuous energy input of its constituting active units~\cite{marchetti2013hydrodynamics,bechinger2016}.
These units, for example, self-propelling agents, are frequently embedded in a surrounding medium. For instance, we think of active suspensions~\cite{lauga2009hydrodynamics,elgeti2015physics} or biofilms \cite{hall2004bacterial}. In these abundant situations, the interplay between activity and the material properties of the enclosing medium is key in determining the overall, collective spatiotemporal dynamics of the system.
Considering the long-term flow behavior, we distinguish between two primary idealized types: active fluids and active solids.

Active fluids are characterized by their capability of terminal flow. Releasing stresses after longer times of displacement from the original positions leads to persistent finite displacements of the volume elements. Restoring forces that take the volume elements the full way back to their initial positions do not exist.
Spontaneously emergent flows in active fluids are directly induced by the enclosed active agents.
Primary examples include active nematics~\cite{doostmohammadi2018active} and suspensions of microswimmers~\cite{lauga2009hydrodynamics,elgeti2015physics} such as bacteria. They exhibit a rich variety of dynamic spatiotemporal patterns~\cite{be2019statistical,aranson2022bacterial,nishiguchi2018engineering,reinken2024pattern}.
Apart from complex vortex and swirling states, such as active turbulence~\cite{dombrowski2004self,wensink2012meso,reinken2018derivation,alert2021active}, one of the most important phenomena is flocking or swarming~\cite{jeckel2019learning,be2020phase}.
These states of polar orientational order and directed collective motion are already found for so-called `dry' active matter. They are described by the seminal Vicsek model~\cite{vicsek1995novel} and its continuum counterpart, the Toner--Tu theory~\cite{toner1998flocks,toner2005hydrodynamics}.

In contrast to active fluids, active solids feature reversible elastic deformations instead of terminal viscous dissipative flows. Due to their solidity, they are unable to support continuous directed distortional motion.
In schematic agent-based descriptions, active solids are frequently represented by active elastic spring networks. That is, the nodes connecting the springs are forced to propel, for example, due to active agents~\cite{shen2016probing,ferrante2013elasticity,hernandez2024model,baconnier2024noise}. Accordingly, a continuous embedding solid-like medium is considered to be discretized into individual elastic springs. 
Other types of active solids are based on contractile or extensile active elements that drive deformations by input of active stresses~\cite{hawkins2014stress,maitra2019oriented}. A prime example of biological active solids is muscular tissue~\cite{needleman2017active}. 
Phenomena of interest in active solids include the nature of their excitations~\cite{baconnier2022selective,caprini2023entropons,kinoshita2025collective} and how polar orientational order develops and spreads~\cite{laang2024topology}.
Furthermore, the phenomenon of odd elasticity has sparked a significant amount of interest in recent years~\cite{scheibner2020odd,fruchart2023odd}.

Numerous existing continuum theories focus on these two limits of active fluids with simple~\cite{toner2005hydrodynamics,saintillan2013active,reinken2018derivation} or complex rheology~\cite{hemingway2015active,hemingway2016viscoelastic,plan2020active,reinken2024vortex,reinken2025self} and active solids~\cite{kopf2013non,maitra2019oriented,scheibner2020odd}. 
Yet, many living biological media do not fit neatly into only one of these two categories of either active viscous fluids or active elastic solids. 
So far, a number of studies on active fluids have extended considerations on collective motion 
to more complex rheology. Examples comprise viscoelasticity~\cite{hemingway2015active,hemingway2016viscoelastic,plan2020active,choudhary2023orientational} or shear thickening and thinning~\cite{mathijssen2016upstream,reinken2024vortex,reinken2025self}.
In particular, the interplay between viscoelasticity and activity in fluid-like systems, concerning their long-term behavior, has been explored in the past to some extent, focusing, for example, on nematic systems~\cite{hemingway2015active,hemingway2016viscoelastic}, multicomponent gels~\cite{joanny2007hydrodynamic}, and the motion of probe particles~\cite{duclut2024probe}.

Correspondingly, several synthetic and many biological materials exhibit both viscous and elastic properties at the same time~\cite{li2021microswimming}. 
For instance, biofilms feature complex rheological behavior~\cite{jana2020nonlinear}. Bacterial biofilms are formed when bacteria excrete extracellular polymeric substances so that they are enclosed by the polymeric material~\cite{hall2004bacterial,worlitzer2022biophysical}.
Depending on the stage of formation of the biofilm, the location within it, and the time scales considered, the material may rather resemble an active fluid or an active solid, but generally show a viscoelastic stage in between. These active biological media thus develop their character from more fluid-like to more solid-like behavior over time or over space. %

\begin{figure}
\includegraphics[width=0.9\linewidth]{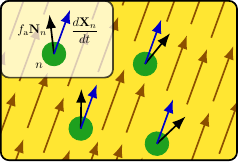}
\caption{\label{fig:MicroscopicPicture}Sketch of the microscopic picture that we use for coarse-graining to derive our continuum theory. $N$ active agents, labeled by $n$, are embedded in a surrounding continuous viscous, viscoelastic, or elastic medium, indicated by the yellow background.
Every active agent $n$ exerts an active force $f_\mathrm{a}\mathbf{N}_n$ (black arrows) on the surrounding medium, see the inset. This force results from self-propulsion by interaction with an underlying substrate or due to external forcing in artificial systems. $f_\mathrm{a}$ is the magnitude of the active force and $\mathbf{N}_n$ its direction, here identified with the orientation of the $n$th object. 
Flows $\mathbf{v}$ (brown arrows) in the medium affect the orientations $\mathbf{N}_n$ of the active objects and lead to their displacements, resulting in velocities $d\mathbf{X}_n/dt$ (blue arrows).}
\end{figure}

As a consequence, there is a need for a corresponding unified framework that is capable of characterizing, simply by variations of parameter values, a full range of viscoelastic behavior, from the limit of active fluid- to the limit of active solid-like.
A comprehensive description combining activity, entering via the self-propulsion of active agents, and viscoelastic material behavior that can be continuously tuned from an active viscous fluid to an active elastic solid is to be presented.

The aim of our work is to derive such a unified framework. For this purpose, we start from a microscopic picture of active objects self-propelling on a substrate and enclosed by a viscoelastic medium. The underlying microscopic equations are coarse-grained to a continuum description. 
By variation of a single parameter, our resulting equations can be tuned continuously from the limit of active fluid-like to the limit of active solid-like behavior, covering a broad range of viscoelasticity in between. Additionally, we explore spatially uniform analytical solutions of the derived equations. In this way, we identify a disordered isotropic state for low activity. An orientationally ordered, polar state emerges at elevated activity but weak elasticity. Moreover, a dynamic state of rotating polar orientational order is found for both stronger activity and elasticity.
Analytical calculations show how the transition from stationary to rotating states changes from supercritical to subcritical with increasing elasticity. 
Our analytical results apply to systems of small size under periodic boundary conditions, where the solutions remain spatially uniform. 
The more general case is referred to numerically in a companion article~\cite{reinken2025rheologically} and demonstrates spatial decorrelation for larger systems.

\section{Microscopic picture}
\label{sec:MicroscopicModel}

Our central goal in the following is to derive our description from a basic, generic microscopic picture, see Fig.~\ref{fig:MicroscopicPicture}. 
For this purpose, we start from an ensemble of $N$ self-propelled objects suspended in a continuous medium.
This framework includes the concept of microswimmers in fluid-like environments and active agents in elastic materials.
The position $\mathbf{X}_n$ and orientation $\mathbf{N}_n$ of object $n$ vary over time $t$ according to the following overdamped Langevin equations,
\begin{align}
\label{eq:LangevinPosition}
\frac{d \mathbf{X}_n}{dt} &= \mathbf{v} + \sqrt{2 D_\mathrm{tr}} \boldsymbol{\xi}_n\, , \\
\label{eq:LangevinOrientation}
\frac{d \mathbf{N}_n}{dt} &=  \boldsymbol{\Pi}(\mathbf{N}_n) \cdot \big[ (\boldsymbol{\Omega} + \tilde{\kappa} \boldsymbol{\Sigma}) \cdot \mathbf{N}_n + \gamma_\mathrm{a} \mathbf{v} + \sqrt{D_\mathrm{r}}\boldsymbol{\eta}_n \big]\, .
\end{align}
In these equations, $\mathbf{v}$, $\boldsymbol{\Omega}$, and $\boldsymbol{\Sigma}$ are  fields that depend on the position $\mathbf{x}$ within the system and time $t$. $\mathbf{v}$ denotes the locally averaged total velocity field of the suspension at position $\mathbf{x}$ and time $t$, which will be discussed in detail in Sec.~\ref{sec:HydrodynamicsAndElasticity}.

The first equation describes advective transport of the active objects with velocity $\mathbf{v}$ and additionally includes a Gaussian white noise $\boldsymbol{\xi}_n$ with a translational diffusion coefficient $D_\mathrm{tr}$.
Activity enters via the forces that the self-propelled objects exert on the surrounding medium to set it and themselves into motion, see below. %

Equation~(\ref{eq:LangevinOrientation}) quantifies the dynamics of the unit vector $\mathbf{N}_n$ that encodes the current orientation of object $n$.
Within the square brackets, the first contribution %
describes reorientations by the local vorticity of the flow field, $\boldsymbol{\Omega} = [(\nabla \mathbf{v})^\top - (\nabla \mathbf{v})]/2$~\cite{pedley1992hydrodynamic}. Similar effects result from distortional gradients of the flow as included by the deformation rate tensor $\boldsymbol{\Sigma} = [(\nabla \mathbf{v})^\top + (\nabla \mathbf{v})]/2$~\cite{pedley1992hydrodynamic}.
The parameter $\tilde{\kappa}$ denotes the tumbling parameter and is a function of the ratio $r$ of major to minor axis of the body of the active object. More precisely, $
\tilde{\kappa} = (r^2 - 1)/(r^2 + 1)$.
Thus, $\tilde{\kappa} = 0$ for spherical and $\tilde{\kappa} >0$ for elongated objects~\cite{pedley1992hydrodynamic}.
Next, the term $\gamma \mathbf{v}_a$ describes alignment with the direction of the overall flow.
For this effect to emerge, the self-propelled units need to be able to ``hold on'' to something to withstand pure advection by the flow, such as a supporting substrate in thin active films~\cite{brotto2013hydrodynamics,maitra2020swimmer,liu2021viscoelastic}. 
Assuming that the active agents exhibit anisotropic friction with respect to the enclosing medium and are capable of reorienting in flows, movement relative to the medium induces rotations~\cite{dadhichi2018origins}.
Specifically, active agents experiencing higher friction at the rear than at the front reorient into the direction of the flow, and vice versa.
Our description takes into account this ``weathervane'' effect already on the microscopic level.
We remark that a similar term is included in various discretized models of active solids~\cite{szabo2006phase,lam2015self,baconnier2022selective,baconnier2025self}.
Equation~(\ref{eq:LangevinOrientation}) adds a Gaussian white noise term $\boldsymbol{\eta}_n$ to the current orientation with rotational diffusion constant $\textbf{$D_\mathrm{r}$}$.
Finally, the overall equation is multiplied by the projector $\boldsymbol{\Pi}(\mathbf{N}_n) = \mathbf{I} - \mathbf{N}_n \mathbf{N}_n$, where $\mathbf{I}$ refers to the unit matrix. This projection maintains the nature of $\mathbf{N}_n$ as a unit vector.

Our study focuses on systems in which the interactions between the self-propelled objects are primarily mediated by the continuous medium in between.
This seems justified for low to intermediate concentrations of active units, that is, when these units rarely approach closely to each other. For higher concentrations, additional interactions, such as steric repulsion, should be considered as well.

\section{Polar orientational order parameter}

We now perform our coarse-graining procedure. For this purpose, we first define the one-object probability distribution function $\mathcal{P}(\mathbf{x},\mathbf{n},t)$ via
\begin{equation}
\label{eq:ProbabilityDensityDistributionFunction}
\mathcal{P}(\mathbf{x},\mathbf{n},t) = \frac{1}{N} \sum_{n=1}^N \bigg\langle \delta(\mathbf{x} - \mathbf{X}_n(t))\delta(\mathbf{n} - \mathbf{N}_n(t))   \bigg\rangle\, .
\end{equation}
It quantifies the probability to find a self-propelled object at position $\mathbf{x}$ with orientation $\mathbf{n}$ at time $t$.  $\mathbf{n}$ is another unit vector. 

Applying standard techniques~\cite{risken1984fokker}, we obtain the Fokker--Planck equation
\begin{align}
\label{eq:Fokker--Planck}
\partial_t \mathcal{P}(\mathbf{x},\mathbf{n},t) = &-\nabla \cdot [\mathbf{v}\mathcal{P}(\mathbf{x},\mathbf{n},t)] + D_\mathrm{tr}\nabla^2\mathcal{P}(\mathbf{x},\mathbf{n},t) \nonumber \\ 
&- \nabla_\mathbf{n} \cdot \big\{ \big[ \boldsymbol{\Omega} \cdot \mathbf{n} + \tilde{\kappa} \boldsymbol{\Pi}(\mathbf{n}) \cdot \boldsymbol{\Sigma} \cdot \mathbf{n}\nonumber \\ 
&\qquad + \gamma_\mathrm{a} \boldsymbol{\Pi}(\mathbf{n}) \cdot \mathbf{v} - D_\mathrm{r} \mathbf{n} \big]\mathcal{P}(\mathbf{x},\mathbf{n},t) \big\}\nonumber \\ 
&+ \frac{1}{2}D_\mathrm{r} \nabla_\mathbf{n}\nabla_\mathbf{n} : [\boldsymbol{\Pi}(\mathbf{n}) \mathcal{P}(\mathbf{x},\mathbf{n},t)]\, .
\end{align}
Here, the first two terms %
describe translational drift and translational diffusion, respectively.
The next term emerges due to rotational drift under the influence of the flow $\mathbf{v}$ of the surrounding medium.
Rotational diffusion causes the contribution by the last term.
Again, we use the abbreviation $\boldsymbol{\Pi}(\mathbf{n}) = \mathbf{I} - \mathbf{n}\mathbf{n}$.

We define the marginal positional probability density $\mathcal{P}(\mathbf{x},t)$ and the concentration field of active objects $c(\mathbf{x},t)$ via
\begin{equation}
c(\mathbf{x},t) = N \mathcal{P}(\mathbf{x},t) = N \int \mathcal{P}(\mathbf{x},\mathbf{n},t) \, d\mathbf{n}\, .
\end{equation}
Next, we obtain dynamic equations for orientational order parameter fields. We concentrate on the polar order vector field $\mathbf{P}$ and the nematic order parameter tensor field $\mathbf{Q}$. %
These order parameter fields are directly connected to the orientational moments $\langle \mathbf{n} \rangle$ and $\langle \mathbf{n}\mathbf{n}\rangle$, respectively. %
That is, 
\begin{equation}
\mathbf{P}(\mathbf{x},t) = \langle \mathbf{n}\rangle(\mathbf{x},t)\, .
\end{equation}
and
\begin{equation}
\label{eq:NematicOrderParameter}
\mathbf{Q}(\mathbf{x},t) = \langle \mathbf{n}\mathbf{n}\rangle^\mathrm{ST} = \langle \mathbf{n} \mathbf{n}\rangle - \mathbf{I}/3\, ,
\end{equation}
where $(\dots)^\mathrm{ST}$ denotes the symmetric traceless part of a tensor.
Here and below, we need the first three orientational moments. They are calculated as the averages over the fluctuating microscopic degrees of freedom via
\begin{equation}
\label{eq:OrientationalMoments}
\begin{aligned}
\langle \mathbf{n} \rangle(\mathbf{x},t) &= \frac{N}{c(\mathbf{x},t)}\int \mathcal{P}(\mathbf{x},\mathbf{n},t) \mathbf{n} \, d\mathbf{n}\, , \\
\langle \mathbf{n} \mathbf{n}  \rangle(\mathbf{x},t) &=\frac{N}{c(\mathbf{x},t)}\int \mathcal{P}(\mathbf{x},\mathbf{n},t)\mathbf{n}\mathbf{n} \, d\mathbf{n} \, , \\
\langle \mathbf{n} \mathbf{n}  \mathbf{n}  \rangle(\mathbf{x},t) &= \frac{N}{c(\mathbf{x},t)}\int \mathcal{P}(\mathbf{x},\mathbf{n},t) \mathbf{n}\mathbf{n}\mathbf{n}\, d\mathbf{n}\, .
\end{aligned}
\end{equation}

Based on the definitions above, we obtain the evolution equations for the concentration and the orientational order parameter fields from the Fokker--Planck equation via projection.
We first determine the equation for the concentration field $c$ of active objects from Eq.~(\ref{eq:Fokker--Planck}) through multiplication by $N$ and integration over all orientations, 
\begin{equation}\label{eq:c}
\partial_t c = - \nabla \cdot (c \mathbf{v}) + D_\mathrm{tr}\nabla^2 c \, .
\end{equation}
The first term on the right-hand side describes advection, that is, transport of concentration with the local velocity field of the suspension.
Assuming incompressibility everywhere in the system, the total velocity field is divergence-free, $\nabla \cdot \mathbf{v} = 0$. 
Moreover, the right-hand side contains a diffusive term. It smoothens variations in concentration to spatially homogeneous distributions of the concentration field. 
Overall, an initially uniform concentration field remains homogeneous. Equation~(\ref{eq:c}) does not contain any contributions that would lead to an accumulation in the concentration of active objects when starting from a homogeneous initial distribution.
Therefore, in the following, we neglect the impact of variations in concentration.

An equation for the field $\mathbf{P}$ of polar orientational order is obtained from the Fokker--Planck equation, Eq.~(\ref{eq:Fokker--Planck}), via multiplication by $N\mathbf{n}$, integration over all orientations, and division by the concentration $c$.
This procedure yields
\begin{eqnarray}
\partial_t \mathbf{P} &= &{}-\nabla \cdot \langle \mathbf{v}\mathbf{n}\rangle + D_\mathrm{tr} \nabla^2 \mathbf{P} - D_\mathrm{r} \mathbf{P} + \langle \boldsymbol{\Omega} \cdot \mathbf{n} \rangle \nonumber\\
&&{}+ \tilde{\kappa} \langle (\mathbf{I} - \mathbf{n}\mathbf{n})\cdot \boldsymbol{\Sigma} \cdot \mathbf{n} \rangle + \gamma_\mathrm{a} \langle (\mathbf{I} - \mathbf{n}\mathbf{n})\cdot \mathbf{v} \rangle \, .\quad
\label{eq:EvolutionPolarOrderStep1}
\end{eqnarray}
Here, the velocity field $\mathbf{v}$ of the suspension is considered as an already locally averaged quantity. Consequently, we take it out of the averages in Eq.~(\ref{eq:EvolutionPolarOrderStep1}).
The same argument holds for the vorticity and deformation rate tensors. Thus, Eq.~(\ref{eq:EvolutionPolarOrderStep1}) yields
\begin{align}
\label{eq:EvolutionPolarOrderStep2}
\partial_t \mathbf{P} = &- \mathbf{v} \cdot \nabla \mathbf{P}  + D_\mathrm{tr} \nabla^2 \mathbf{P} - D_\mathrm{r} \mathbf{P} + \boldsymbol{\Omega} \cdot \mathbf{P} \\
&+ \tilde{\kappa} \boldsymbol{\Sigma} \cdot \mathbf{P}- \tilde{\kappa}  \boldsymbol{\Sigma} : \langle \mathbf{n}\mathbf{n}\mathbf{n} \rangle  + \gamma_\mathrm{a} \mathbf{v} - \gamma_\mathrm{a}  \mathbf{v} \cdot \langle \mathbf{n}\mathbf{n}\rangle \, . \nonumber
\end{align}

This equation contains the higher-order moments $\langle \mathbf{n}\mathbf{n} \rangle$ and $\langle \mathbf{n}\mathbf{n}\mathbf{n} \rangle$.
As discussed above, the second moment is connected to the nematic order parameter tensor via Eq.~(\ref{eq:NematicOrderParameter}).
Since we mostly consider polar objects without head-tail symmetry, we do not address the equation for the second moment explicitly, but rather employ closure relations for higher-order moments.
Specifically, we use a standard quadratic closure approximation~\cite{doi1988theory,rienaecker1998orientational,kroeger2008consistent} to write the nematic order parameter tensor $\mathbf{Q}$ as a function of the polar order parameter vector $\mathbf{P}$ via
\begin{equation}
\label{eq:QuadraticClosure}
\mathbf{Q} = (\mathbf{P}\mathbf{P})^\mathrm{ST} = \mathbf{P}\mathbf{P} - (\mathbf{P}\cdot\mathbf{P})\mathbf{I}/3\, .
\end{equation}
Combining this expression with the definition of the nematic order parameter tensor, Eq.~(\ref{eq:NematicOrderParameter}), yields for the second moment
\begin{equation}
\label{eq:SecondMomentClosure}
\langle \mathbf{n}\mathbf{n} \rangle = \mathbf{Q} +  \mathbf{I}/3 = \mathbf{P}\mathbf{P} + (1 - \mathbf{P}\cdot\mathbf{P})\mathbf{I}/3\, .
\end{equation}

The third moment is treated via the so-called Hand-closure~\cite{hand1962theory,kroeger2008consistent}, see Ref.~\onlinecite{reinken2018derivation} for a more detailed discussion. It implies
\begin{equation}
\label{eq:HandClosure}
\langle \mathbf{n}\mathbf{n}\mathbf{n}\rangle^\mathrm{ST} = \mathbf{0}\, .
\end{equation}
The components of the symmetric traceless part are calculated from the full tensor via
\begin{equation}
\label{eq:SymmetricTracelessThirdOrderTensor}
\langle \mathbf{n}\mathbf{n}\mathbf{n}\rangle^\mathrm{ST}_{ijk} = \langle n_i n_j n_k \rangle - \frac{1}{5}(\delta_{ij} \langle n_k\rangle  + \delta_{ki} \langle n_j\rangle + \delta_{jk} \langle n_i\rangle ) \, .
\end{equation}
Employing the Hand closure, the right-hand side of Eq.~(\ref{eq:SymmetricTracelessThirdOrderTensor}) must be zero. 
We thus obtain the third moment as a function of the polar order parameter vector field $\mathbf{P}$.
This yields the components
\begin{equation}
\label{eq:ThirdMomentClosure}
\langle n_i n_j n_k \rangle = \frac{1}{5}(\delta_{ij} P_k + \delta_{ki} P_j + \delta_{jk} P_i) \,.
\end{equation}
Employing both Eqs.~(\ref{eq:SecondMomentClosure}) and (\ref{eq:ThirdMomentClosure}), we find from Eq.~(\ref{eq:EvolutionPolarOrderStep2})
\begin{align}
\label{eq:EvolutionPolarOrderStep3}
\partial_t \mathbf{P} = &- \mathbf{v} \cdot \nabla \mathbf{P} + \boldsymbol{\Omega} \cdot \mathbf{P} + \kappa \boldsymbol{\Sigma} \cdot \mathbf{P}  + D_\mathrm{tr} \nabla^2 \mathbf{P} - D_\mathrm{r} \mathbf{P} \nonumber \\
&+ \gamma_\mathrm{a} \mathbf{v} \cdot [(2 + \mathbf{P}\cdot\mathbf{P})\mathbf{I}/3 - \mathbf{P}\mathbf{P}] \, ,
\end{align}
where $\kappa = 3 \tilde{\kappa}/5$.
The nonlinear term with coefficient $\gamma_\mathrm{a}$ is a consequence of the closure approximation for the nematic order parameter tensor field $\mathbf{Q}$ as discussed above. It effectively restricts the polar order parameter field to values $|\mathbf{P}| \leq 1$.
Equation~(\ref{eq:EvolutionPolarOrderStep3}) corresponds to our final form of the evolution equation for the polar order parameter vector field $\mathbf{P}$.

\section{Hydrodynamics and elasticity}
\label{sec:HydrodynamicsAndElasticity}

Having addressed the orientational order parameter fields, we next turn to the coupling between Eq.~(\ref{eq:EvolutionPolarOrderStep3}) and the overall velocity field $\mathbf{v}$ of the suspension as well as the degrees of freedom associated with elasticity. The systematic path of introducing further degrees of freedom, particularly those associated with spontaneous symmetry breaking, into the theory of hydrodynamics was outlined more than half a century ago \cite{martin1972unified}. Specifically, this concerns broken translational symmetry in elastic solids, associated with an additional field of 
displacements $\mathbf{u}$. Its extension to viscoelastic systems was systematically derived within this framework more than a quarter of a century ago \cite{temmen2000convective}. Here, we follow this path of generalized hydrodynamics and employ it to analyze the properties of active viscoelastic suspensions on substrates.

It is the Euler point of view that is associated with generalized hydrodynamics. This perspective allows for the interpretation within the hydrodynamics framework of what is usually referred to as displacement field $\mathbf{u}$ in elasticity theory. Introducing elasticity into generalized hydrodynamics, $\mathbf{u}$ can be viewed as the field of currently and locally ``memorized'' elastic displacements in the system \cite{puljiz2019memory}.

Usually, one starts the list of hydrodynamic equations by the continuity equation, that is, the dynamic equation for the density. We consider the active viscoelastic material
as incompressible with approximately constant density.
For biological and particularly aqueous systems, this frequently represents a reasonable assumption. Thus, the hydrodynamic continuity equation simplifies to $\nabla \cdot \mathbf{v} = 0$.

Next, we turn to the dynamic equation for the flow velocity itself. For simple liquids, it is given by the Navier-Stokes equation. We here consider low-Reynolds number conditions and overdamped dynamics. Additional linear friction with a supporting substrate is included. Therefore, the inertial terms are neglected, and we evaluate an amended Stokes equation,
\begin{equation}
\label{eq:StokesEquation}
\mathbf{0} = {}- \nabla p + \eta \nabla^2 \mathbf{v} + \mu \nabla^2 \mathbf{u} - \nu_\mathrm{v} \mathbf{v} - \nu_\mathrm{d} \mathbf{u} + \mathbf{f}\, .
\end{equation}
It describes the balance of multiple forces, which we explain in the following.

On the right-hand side, for incompressible systems, the pressure field $p$ effectively acts as a Lagrange multiplier enforcing incompressibility $\nabla\cdot\mathbf{v}=0$.
The second term introduces the viscous forces quantified by the shear viscosity $\eta$, as usually in the Stokes equation. This contribution is developed from the divergence of the stress tensor, $\nabla\cdot\bm{\sigma}$, for isotropic, incompressible fluids \cite{landau1987fluid}. 
Yet, if the system shows additional elastic features, also elastic stresses $\bm{\sigma}^{\mathrm{el}}$ contribute to the overall hydrodynamic stress. For isotropic, incompressible systems $\bm{\sigma}^{\mathrm{el}}=2\mu\mathbf{U}$, where $\mathbf{U}$ denotes the strain tensor. In linear elasticity, it is typically expressed in terms of the elastic displacement field $\mathbf{u}(\mathbf{x},t)$.
The latter includes the reversible elastic displacements stored (``memorized'') at time $t$ for the material elements currently located at positions $\mathbf{x}$. Moreover, in linear elasticity theory
$\mathbf{U}=\left(\nabla\mathbf{u}+(\nabla\mathbf{u})^{\mathrm{T}}\right)/2$ \cite{landau1986theory,  chaikin1995principles, pleiner2004nonlinear}, while incompressible systems satisfy $\nabla\cdot\mathbf{u}=0$ \cite{landau1986theory}. Together, we recover from $\nabla\cdot\bm{\sigma}^{\mathrm{el}}$ the third term on the right-hand side of Eq.~(\ref{eq:StokesEquation}). 
The shear modulus $\mu$ is a measure for the mechanical stiffness of the material, that is, its resistance against elastic deformations.

Since we consider thin films of active suspensions on substrates, additional frictional forces linear in overall velocity $\mathbf{v}$ are taken into account. $\nu_\mathrm{v}$ is the corresponding friction coefficient.
We further allow for elastic anchoring or pinning to the supporting substrate, which leads to elastic restoring forces. They tend to drive the material elements back to elastically memorized positions as quantified by the field $\mathbf{u}$.
We here consider corresponding restoring forces to be linear as well and set their strength by the parameter $\nu_\mathrm{d}$.

Finally, all additional forces acting on the material are summarized by the force density field $\mathbf{f}$. In our case, these additional forces are mainly due to the activity of the active objects.
In the simplest case, the $n$th active object exchanges momentum with the substrate and thus can exert an active force $\mathbf{f}_n$ along its polar orientation $\mathbf{N}_n$ on the active film, $\mathbf{f}_n = f_a \mathbf{N}_n$, where $f_a$ is the magnitude of this force.
Upon averaging, the force density field $\mathbf{f}$ in Eq.~(\ref{eq:StokesEquation}) becomes
\begin{equation}
\label{eq:AciveForcing}
\mathbf{f} = c f_a \langle \mathbf{n} \rangle = \nu_\mathrm{p} \mathbf{P}\, .
\end{equation}
We introduced the strength of active forcing, $\nu_\mathrm{p} = c f_a$.
This kind of contribution corresponds to propulsive active forces often employed when modeling active solids~\cite{maitra2019oriented,ferrante2013elasticity,xu2023autonomous,caprini2023entropons}.

The net force density %
results from the interaction with the substrate. Otherwise, the active units would be force-free. That is, they would not be able to exert net forces but, in the lowest-order case, net stresses on their enclosing environment.
Such net force-free contributions are often assumed in descriptions of bulk active fluids, leading to active stresses. %
They have been taken into account before%
~\cite{simha2002hydrodynamic,saintillan2008instabilities,reinken2018derivation}.

In the systematic approach of generalized hydrodynamics, each variable associated with a spontaneously broken symmetry requires and follows a separate dynamic equation as well \cite{martin1972unified}. Integrating elasticity into generalized hydrodynamics, this spontaneously broken symmetry is reflected by elastic displacements. In Refs.~\onlinecite{temmen2000convective, pleiner2004nonlinear}, the associated dynamic equation was formulated in terms of the elastic strain tensor $\mathbf{U}$. The correspondence between this tensorial approach and a formulation in terms of a vectorial quantity was pointed out \cite{temmen2000convective}, see also Appendix~\ref{app:DisplacementFieldDynamics}.

Using again the identification $\mathbf{U}=\left(\nabla\mathbf{u}+(\nabla\mathbf{u})^{\mathrm{T}}\right)/2$, the systematically derived basic dynamic equation resulting within the framework of generalized hydrodnamics for $\mathbf{U}$ \cite{pleiner2004nonlinear} can be reduced to the vectorial dynamic equation
\begin{equation}
\label{eq:EvolutionEquationDisplacement}
\partial_t \mathbf{u} + \mathbf{v} \cdot \nabla \mathbf{u}  = \mathbf{v} - \tau_\mathrm{d}^{-1} \mathbf{u} - \nabla q\, .
\end{equation}
In brief, full linearization leads to $\partial_t \mathbf{u}  = \mathbf{v} - \tau_\mathrm{d}^{-1} \mathbf{u}$, which was presented before \cite{puljiz2019memory, richter2021rotating}. In this expression, generation of elastic, memorized displacements $\mathbf{u}$ through motion of material elements $\mathbf{v}$ is obvious.  
At the same time, in viscoelastic substances internal stress and strain or, correspondingly, memorized elastic displacements, may decay over time. This contribution was outlined in the generalized hydrodynamic theory as well \cite{temmen2000convective, pleiner2004nonlinear}. For example, strained entangled polymeric melts may relax their stress and strain through disentanglement processes \cite{de1979scaling, doi1988theory}. 
The decay rate of this elastic memory is set by the inverse relaxation time $\tau_\mathrm{d}$.

Furthermore, in Eq.~(\ref{eq:EvolutionEquationDisplacement}), we perform one step beyond the completely linearized picture. We additionally keep, during linearization of the nonlinear description of generalized hydrodynamics \cite{temmen2000convective, pleiner2004nonlinear}, the term of convective transport $\mathbf{v}\cdot\nabla\mathbf{u}$. Still, the equation remains linear in $\mathbf{u}$.
Such transport becomes important when considering viscoelastic fluids that are able to support persistent flow.
On the right-hand side of Eq.~(\ref{eq:EvolutionEquationDisplacement}), it requires the additional contribution $-\nabla q$ that emerges when consistently maintaining incompressibility $\nabla \cdot \mathbf{u} = 0$ of the system when we keep the description linear in $\mathbf{u}$.
Further details on Eq.~(\ref{eq:EvolutionEquationDisplacement}) together with associated assumptions are summarized in Appendix~\ref{app:DisplacementFieldDynamics}. 

In this context, we review once more 
the interpretation of
the field of elastic displacements $\mathbf{u}$ in the generalized hydrodynamics framework. We recall that the generalized hydrodynamics approach is genuinely field-based, not trajectory-based. In other words, it takes the Euler perspective, instead of the Lagrange point of view \cite{chaikin1995principles, temmen2000convective, menzel2025linear}. The Euler perspective quantifies the material in its current state at time $t$, at each position $\mathbf{x}$ in space. Accordingly, the field of elastic displacements $\mathbf{u}(\mathbf{x},t)$ does not describe to which position a material element moves from position $\mathbf{x}$. Instead, it quantifies from which position $\mathbf{x}-\mathbf{u}(\mathbf{x},t)$ the material element presently located at position $\mathbf{x}$ has 
come from.
Correspondingly, Ref.~\onlinecite{temmen2000convective} in the generalized hydrodynamics framework bases the theory on an initial field $\bm{a}(\mathbf{x},t)=\mathbf{x}-\mathbf{u}(\mathbf{x},t)$. In this sense, $\mathbf{u}$ encodes the memorized positions, from where the material elements have moved to $\mathbf{x}$. To express these relations and emphasize the Euler perspective, we refer to $\mathbf{u}$ as the elastic memory field or field of memorized elastic displacements.
Thus, if all stresses were released at time $t$, the material elements would tend to return to positions $\mathbf{x}-\mathbf{u}(\mathbf{x},t)$. 

This ``retrospective'' Euler point of view leads to marked differences of interpretation of the quantities when compared to Lagrange descriptions. Specifically, this becomes apparent in the context of viscoelasticity. Above, we mentioned the relaxation of entangled polymer melts when stretched into a strained state and maintained in this situation. Elastic stress and strain are directly related to each other in linear elasticity for isotropic, incompressible materials via $\bm{\sigma}^{\mathrm{el}}=2\mu\mathbf{U}=\mu\left(\nabla\mathbf{u}+(\nabla\mathbf{u})^{\mathrm{T}}\right)$, see also Appendix~\ref{app:DisplacementFieldDynamics}. Relaxation of stress and strain in a viscoelastic system therefore implies decay of $\mathbf{u}$. Naturally, the space positions $\mathbf{x}$ of physical space, where $\mathbf{u}(\mathbf{x},t)$ is defined and evaluated, remain unaffected by this process of relaxation. Thus, it is $\mathbf{x}-\mathbf{u}(\mathbf{x},t)$ that changes. These are the memorized positions, where the material elements would displace back to upon release of stresses. In the Euler perspective, the memorized positions approach the current locations $\mathbf{x}$ during viscoelastic relaxation. The material elements presently located at $\mathbf{x}$ would shift back to positions closer to $\mathbf{x}$ than before relaxation. For a basic graphical illustration we refer to a recent review on the generalized hydrodynamics approach for passive systems \cite{menzel2025linear}. In other words, the system forgets about the initial unstrained state. Its elastic memory, quantified by $\mathbf{u}$, decays so that the system relaxes into a new unstrained state. The rate of decay of this memory is $\tau_\mathrm{d}^{-1}$ in Eq.~(\ref{eq:EvolutionEquationDisplacement}).

Varying the relaxation time $\tau_\mathrm{d}$ of the elastic memory, we can continuously tune the behavior of the material from the limit of purely fluid-like to the opposite limit of purely elastic, solid-like.
More precisely, the limit of a viscous fluid is obtained for instant relaxation $\tau_\mathrm{d} \to 0$. 
Contrarily, the limit $\tau_\mathrm{d} \to \infty$ of diverging relaxation time describes a non-decaying elastic memory. In this case, the behavior is perfectly elastic and all displacements are completely reversible.
Still, in this solid-like case, damping by viscous and frictional forces affects the dynamics.
Indeed, Eq.~(\ref{eq:EvolutionEquationDisplacement}) in its linearized version together with the stress-strain relation $\bm{\sigma}^\mathrm{el}=\mu\mathbf{U}$ directly reproduces what is frequently referred to as the basic constitutive linear relations of viscoelasticity. We obtain the Maxwell model for finite $\tau_{\mathrm{d}}$ and $\eta=0$ \cite{pleiner2004nonlinear, menzel2025linear}. For $\tau_\mathrm{d}\rightarrow\infty$, we find the Kelvin-Voigt model \cite{menzel2025linear}. This also implies that, by adjusting one single parameter $\tau_{\mathrm{d}}$, we can tune the description from viscous fluid-like, via a broad range of 
viscoelastic behavior or imperfect elasticity in between, to genuinely elastic systems.

\section{Rescaled dynamic equations}

To reduce the number of parameters, we rescale time and space using the rotational and translational diffusion coefficients of the active agents.
We thus introduce the rescaled time $\tilde{t} = D_\mathrm{r}t$ and rescaled space $\tilde{\mathbf{r}} = \mathbf{r} / \sqrt{D_\mathrm{tr}/D_\mathrm{r}}$, implying the dimensionless velocity field  $\tilde{\mathbf{v}} = \mathbf{v}/\sqrt{D_\mathrm{tr}D_\mathrm{r}}$.
Moreover, the material parameters are rescaled as 
\begin{eqnarray}
\label{eq:CoefficientsRescaled}
&&\tilde{\gamma}_\mathrm{a} = D_\mathrm{tr}^{1/2}\, D_\mathrm{r}^{-1/2} \, \gamma_\mathrm{a}\, , \quad \tilde{\tau}_\mathrm{d} = D_\mathrm{r}\tau_\mathrm{d}\, , \quad \tilde{\nu}_\mathrm{v} =  D_\mathrm{r}^{-1} \rho^{-1} \nu_\mathrm{v} \, , \nonumber \\
&&\tilde{\nu}_\mathrm{d} = D_\mathrm{r}^{-2} \rho^{-1} \nu_\mathrm{d}  \, ,\ \ \tilde{\nu}_\mathrm{p} = D_\mathrm{r}^{-3/2}D_\mathrm{tr}^{-1/2} \rho^{-1} \nu_\mathrm{p}\, , \nonumber \\
&&\tilde{\eta} = D_\mathrm{tr}^{-1} \rho^{-1}\, \eta\, , \quad \tilde{\mu} = D_\mathrm{r}^{-1} D_\mathrm{tr}^{-1} \rho^{-1}\, \mu\, , 
\end{eqnarray}
where $\rho$ is the constant density of the medium.

As a result, the evolution equation for $\mathbf{P}$, Eq.~(\ref{eq:EvolutionPolarOrderStep3}), takes the form
\begin{equation}
\label{eq:EvolutionEquationPolarOrderRescaled}
\begin{aligned}
\partial_{\tilde{t}} \mathbf{P} + \tilde{\mathbf{v}} \cdot \tilde{\nabla} \mathbf{P} = &\, \tilde{\nabla}^2 \mathbf{P} - \mathbf{P} + \tilde{\boldsymbol{\Omega}} \cdot \mathbf{P} + \kappa \tilde{\boldsymbol{\Sigma}} \cdot \mathbf{P}\\ 
&+ \tilde{\gamma}_\mathrm{a} \tilde{\mathbf{v}} \cdot [(2 + \mathbf{P}\cdot\mathbf{P})\mathbf{I}/3 - \mathbf{P}\mathbf{P}] \, .
\end{aligned}
\end{equation}
In terms of the dimensionless elastic memory field $\tilde{\mathbf{u}} = \mathbf{u}/\sqrt{D_\mathrm{tr}/D_\mathrm{r}}$, the evolution equation for the velocity field, Eq.~(\ref{eq:StokesEquation}), becomes
\begin{equation}
\label{eq:StokesEquationRescaled}
0 = - \tilde{\nabla} \tilde{p} + \tilde{\eta} \tilde{\nabla}^2 \tilde{\mathbf{v}}  + \tilde{\mu}\tilde{\nabla}^2 \tilde{\mathbf{u}} - \tilde{\nu}_\mathrm{v} \tilde{\mathbf{v}} - \tilde{\nu}_\mathrm{d} \tilde{\mathbf{u}} + \tilde{\nu}_\mathrm{p} \mathbf{P} \, .
\end{equation}
The rescaled pressure $\tilde{p} = D_\mathrm{r}^{-1} D_\mathrm{tr}^{-1} \rho^{-1} p$ acts as a Lagrange multiplier ensuring incompressibility via a divergence-free velocity field, $\tilde{\nabla} \cdot \tilde{\mathbf{v}} = 0$.
Finally, we recast Eq.~(\ref{eq:EvolutionEquationDisplacement}) for the evolution of the elastic memory field into the dimensionless form
\begin{equation}
\label{eq:EvolutionEquationDisplacementRescaled}
\partial_{\tilde{t}} \tilde{\mathbf{u}} + \tilde{\mathbf{v}} \cdot \tilde{\nabla} \tilde{\mathbf{u}}  = \tilde{\mathbf{v}} - \tilde{\tau}_\mathrm{d}^{-1} \tilde{\mathbf{u}} - \tilde{\nabla} \tilde{q} \, ,
\end{equation}
where $\tilde{q} = D_\mathrm{tr}^{-1} q$.
In the remainder of this work, we evaluate the rescaled equations together with the rescaled material parameters in Eq.~(\ref{eq:CoefficientsRescaled}), omitting the tilde from now on.

\section{Analysis for spatially uniform solutions}
\label{sec:Solutions}

We proceed by an analytical investigation of the resulting dynamics. To be able to perform this task, we determine spatially uniform solutions and set all spatial derivatives to zero. 

We begin with stationary solutions. Thus, we set all time derivatives to zero as well, which yields
\begin{align}
\mathbf{0} &= - \mathbf{P} +\gamma_\mathrm{a} \mathbf{v} \cdot [(2 + |\mathbf{P}|^2)\mathbf{I}/3 - \mathbf{P}\mathbf{P}] \, , \\
\mathbf{0} &= - \nu_\mathrm{v} \mathbf{v} - \nu_\mathrm{d} \mathbf{u} + \nu_\mathrm{p} \mathbf{P} \, , \\
\mathbf{0} &= \mathbf{v} - \tau_\mathrm{d}^{-1} \mathbf{u} \, .
\end{align}
First, the trivial solution,
\begin{equation}
\mathbf{P} = \mathbf{0}\, , \quad \mathbf{v} = \mathbf{0}\, , \quad \mathbf{u} = \mathbf{0}\, ,
\end{equation}
always exists, regardless of the chosen parameter values.
It describes an isotropic state of disordered orientations of the active units, vanishing macroscopic velocity (macroscopic quiescence), and vanishing deformation.

Besides, depending on the parameter values, there exists an additional nontrivial uniform stationary solution.
It spontaneously breaks rotational symmetry. All fields ($\mathbf{P}$, $\mathbf{v}$, $\mathbf{u}$) are spatially homogeneous but nonvanishing.
Without loss of generality, we set $\mathbf{P} = (P_0^\mathrm{st},0)$, $\mathbf{v} = (v_0^\mathrm{st},0)$, and $\mathbf{u} = (u_0^\mathrm{st},0)$, leading to
\begin{equation}
\label{eq:PolarSolution}
\begin{aligned}
P_0^\mathrm{st} &= \sqrt{1 - \frac{3(\nu_\mathrm{v} + \nu_\mathrm{d} \tau_\mathrm{d})}{2 \gamma_\mathrm{a} \nu_\mathrm{p}}}\, , \\
v_0^\mathrm{st} &= \frac{\nu_\mathrm{p} }{\nu_\mathrm{v} + \nu_\mathrm{d} \tau_\mathrm{d}} P_0^\mathrm{st} \, , \\
u_0^\mathrm{st} &= \frac{\nu_\mathrm{p} \tau_\mathrm{d}}{\nu_\mathrm{v} + \nu_\mathrm{d} \tau_\mathrm{d}} P_0^\mathrm{st} \, .
\end{aligned}
\end{equation}
Requiring that $P_0^\mathrm{st}$ is real-valued yields a condition for the existence of this stationary polar dynamic state, which is determined as
\begin{equation}
\label{eq:ConditionPolarStationary}
2 \nu_\mathrm{p} \gamma_\mathrm{a} > 3(\nu_\mathrm{v} + \nu_\mathrm{d} \tau_\mathrm{d})\, .
\end{equation}
Thus, this persistent unidirectional motion requires alignment of the objects with the local velocity, $\gamma_\mathrm{a} > 0$. Moreover, the active forces as quantified by $\nu_\mathrm{p}$, see Eq.~(\ref{eq:AciveForcing}), must be able to overcome the viscous and elastic forces characterized by $\nu_\mathrm{v}$ and $\nu_\mathrm{d} \tau_\mathrm{d}$.

\begin{figure}
\includegraphics[width=0.999\linewidth]{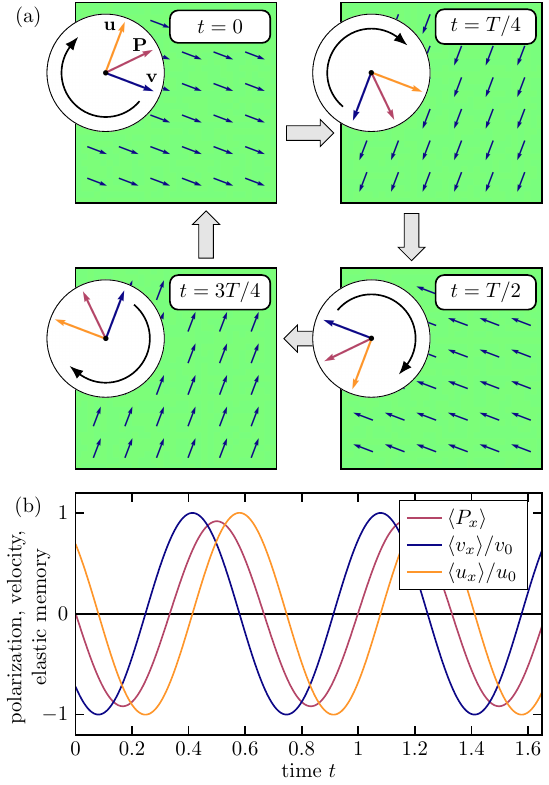}
\caption{\label{fig:RotSol}Spatially uniform dynamic states of rotating polar orientational order and flow. (a) Snapshots of the uniform velocity field during one period $T$ for a genuinely elastic solid ($\tau_\mathrm{d}\rightarrow\infty$). The arrows in the main plots display the current state of the overall flow field $\mathbf{v}$. Circular insets visualize the current states of the polar orientational order parameter $\mathbf{P}$, overall flow velocity $\mathbf{v}$, and field of memorized elastic displacements $\mathbf{u}$. Phase shifts between these quantities are obvious. These results agree with numerical solutions in a small periodic box. (b) Corresponding time evolution of the spatially averaged $x$-components of $\mathbf{P}$, $\mathbf{v}$, and $\mathbf{u}$. We rescale $\langle v_x \rangle$ and $\langle u_x \rangle$ by the amplitudes $v_0$ and $u_0$, respectively. We set the remaining parameters to $\gamma_\mathrm{a} = 1$, $\kappa = 0$, $\eta = 1$, $\mu = 1$, $\nu_\mathrm{v} = 1$, $\nu_\mathrm{d} = 10$, and $\nu_\mathrm{p} = 20$.}
\end{figure}

As a result, when the relaxation time $\tau_\mathrm{d}$ of the memory of displacements is increased, at some point this spatially uniform, stationary, polar solution vanishes.
Particularly, this is the case for elastic solids, for which $\tau_\mathrm{d} \to \infty$.
In this case, intuitively the elastic restoring forces prevent continuous unidirectional motion.

To illustrate the condition in Eq.~(\ref{eq:ConditionPolarStationary}), we may introduce a Deborah number $\mathrm{De}$ in the present context.
This dimensionless number quantifies the ratio between the elastic relaxation time $\tau_\mathrm{d}$ of the viscoelastic medium and some intrinsic time scale $\tau_{a}$ of the considered process, $\mathrm{De} = \tau_\mathrm{d} / \tau_a$.
In our case, the latter is given by the time scale $\tau_a$ of active propulsion, and we define it as $\tau_{a} = (2 \nu_\mathrm{p} \gamma_\mathrm{a} /3 -\nu_\mathrm{v}) /\nu_\mathrm{d} $.
It essentially characterizes
the time scale resulting from actively forced, orientationally ordered motion as reduced by linear friction with the substrate when competing with elastic pinning to the substrate. 
Using these definitions, Eq.~(\ref{eq:ConditionPolarStationary}) reduces to $\mathrm{De}<1$.
Smaller Deborah numbers correspond to a situation in which the viscoelastic medium behaves more fluid-like in the context of the considered process.
In this case, the process of active propulsion is slower than the relaxation of the elastic memory, which enables the emergence of the stationary polar state and persistent migration.
Conversely, for Deborah numbers $\mathrm{De} > 1$, the elastic forces prevent persistent directed motion. Thus, the medium behaves more solid-like.

In addition to the two stationary solutions mentioned above, we find analytical, spatially homogeneous solutions of rotating polar order and flow over time.
Here, we return to Eqs.~(\ref{eq:EvolutionEquationPolarOrderRescaled})%
--(\ref{eq:EvolutionEquationDisplacementRescaled}) and only set all spatial derivatives to zero.
Since Eq.~(\ref{eq:StokesEquationRescaled}) is linear and does not explicitly depend on time, the velocity $\mathbf{v}$ is not an independent quantity.
Furthermore, the incompressibility conditions are fulfilled automatically. The pressure field does not enter the solutions, because we are only considering spatially homogeneous situations. 
The evolution equations thus reduce to
\begin{align}
\label{eq:UniformEquations}
\partial_t \mathbf{P} &= a_P \mathbf{P} - 2 a_u \mathbf{u} - b |\mathbf{P}|^2 \mathbf{P} - a_u |\mathbf{P}|^2 \mathbf{u} + 3a_u (\mathbf{u}\cdot\mathbf{P})\mathbf{P}\, , \nonumber \\
\partial_t \mathbf{u} &= c_P \mathbf{P} - c_u \mathbf{u} \, ,
\end{align}
where we introduced the abbreviations
\begin{equation}
\begin{aligned}
a_P = \frac{2\gamma_\mathrm{a} \nu_\mathrm{p}}{3\nu_\mathrm{v}} - 1\, , \quad a_u &= \frac{\gamma_\mathrm{a}\nu_\mathrm{d}}{3 \nu_\mathrm{v}}\, , \quad b = \frac{2\gamma_\mathrm{a}\nu_\mathrm{p}}{3 \nu_\mathrm{v}}\, , \\\quad c_P = \frac{\nu_\mathrm{p}}{\nu_\mathrm{v}}\, , 
\quad &c_u = \frac{1}{\tau_\mathrm{d}} + \frac{\nu_\mathrm{d}}{\nu_\mathrm{v}}\, .
\end{aligned}
\end{equation}
We write polar orientational order parameter, velocity, and elastic memory fields in polar components, 
\begin{equation}
\label{eq:PolarForm}
\begin{aligned}
\mathbf{P} &= P_0 \! \begin{pmatrix} \cos\phi_P \\ \sin\phi_P \end{pmatrix}\! ,\\ 
\mathbf{v} &= v_0\! \begin{pmatrix} \cos\phi_v \\ \sin\phi_v \end{pmatrix}\! ,\\
\mathbf{u} &= u_0 \! \begin{pmatrix} \cos\phi_u \\ \sin\phi_u \end{pmatrix}\! ,
\end{aligned}
\end{equation}
where $P_0$, $v_0$, and $u_0$ are the magnitudes of $\mathbf{P}$, $\mathbf{v}$, and $\mathbf{u}$, respectively. $\phi_P$, $\phi_v$, and $\phi_u$ quantify the angles of their orientations.
Inserting the expressions for $\mathbf{P}$ and $\mathbf{u}$ into Eqs.~(\ref{eq:UniformEquations}) yields dynamic equations for the amplitudes and angles,
\begin{equation}
\label{eq:UniformEquationsPolar}
\begin{aligned}
\partial_t P_0 = &\ a_P P_0 - b P_0^3 + 2 a_u u_0 P_0^2 \cos(\phi_P - \phi_u)\\ &- 2 a_u u_0 \cos(\phi_P - \phi_u)\, ,\\
\partial_t u_0 = &- c_u u_0 + c_P P_0 \cos(\phi_P - \phi_u)\, ,\\
\partial_t \phi_P = &\frac{a_u u_0 (P_0^2 + 2) }{P_0}\sin(\phi_P - \phi_u)\, ,\\
\partial_t \phi_u = &\frac{c_P P_0}{u_0}\sin(\phi_P - \phi_u)\, .
\end{aligned}
\end{equation}

To continue, we assume persistent rotations of the vectorial quantities associated with phase-shifted harmonic oscillations of their individual components. Thus, the magnitudes have relaxed to a constant value, $\partial_t P_0 = 0$ and $\partial_t u_0 = 0$, and the phases are given by $\phi_P = \omega_0 t$ and $\phi_u = \omega_0 t - \Delta\phi_{Pu}$, where $\Delta\phi_{Pu}$ denotes the phase shift between $\mathbf{P}$ and $\mathbf{u}$.
The resulting set of equations is then solved analytically for the amplitudes $P_0$ and $u_0$, frequency $\omega_0$, as well as the phase shift $\Delta\phi_{Pu}$.

We find multiple solutions, which can be mapped onto one another by switching the sign of the amplitudes, a phase shift, or a simultaneous change in sign of both $\omega_0$ and $\Delta\phi_{Pu}$.
The associated underlying symmetries reflect the process of spontaneous symmetry breaking, which occurs when the system develops one of the solutions.
For simplicity, we here only list the expressions for the dynamic states of positive $P_0$, $u_0$, $\omega_0$, and $\Delta\phi_{Pu}$.

\begin{widetext}
Importantly, there are still two distinct solutions that cannot be mapped onto each other by these transformations. 
We denote them by $+$ and $-$.
Their explicit form is obtained as
\begin{equation}
P_{0\pm} = \sqrt{\frac{6 \nu_\mathrm{v} \tau_\mathrm{d} + 24 \nu_\mathrm{v}- 12 \tau_\mathrm{d} (\gamma_\mathrm{a} \nu_\mathrm{p} - 2 \nu_\mathrm{d}) \pm 2 A}{- 3 \nu_\mathrm{v} \tau_\mathrm{d} + 6 \nu_\mathrm{v}  + 6 \tau_\mathrm{d}  (\gamma_\mathrm{a} \nu_\mathrm{p} + \nu_\mathrm{d}) \mp A}}\, ,
\label{eq:P0pm}
\end{equation}
\begin{equation}
u_{0\pm} = \sqrt{\frac{\nu_\mathrm{p} [3 \nu_\mathrm{v} \tau_\mathrm{d} + 12 \nu_\mathrm{v}  - 6 \tau_\mathrm{d} (\gamma_\mathrm{a} \nu_\mathrm{p} - 2 \nu_\mathrm{d}) \pm A]}{6 \gamma_\mathrm{a} \nu_\mathrm{d}  (\nu_\mathrm{d} \tau_\mathrm{d} + \nu_\mathrm{v})}}\, ,
\end{equation}
\begin{equation}
\omega_{0\pm} = \frac{\sqrt{\nu_\mathrm{d} \tau_\mathrm{d} + \nu_\mathrm{v}}}{\nu_\mathrm{v} \tau_\mathrm{d}}
\sqrt{\frac{12\gamma_\mathrm{a} \nu_\mathrm{p} \nu_\mathrm{d} \tau_\mathrm{d}^2  + [\nu_\mathrm{d} \tau_\mathrm{d} + \nu_\mathrm{v}][3 \nu_\mathrm{v} \tau_\mathrm{d} - 6 \nu_\mathrm{v}  - 6 \tau_\mathrm{d}  (\gamma_\mathrm{a} \nu_\mathrm{p} + \nu_\mathrm{d}) \pm A]}
{-3 \nu_\mathrm{v} \tau_\mathrm{d} + 6 \nu_\mathrm{v}  + 6 \tau_\mathrm{d}  (\gamma_\mathrm{a} \nu_\mathrm{p} + \nu_\mathrm{d}) \mp A}}\, ,
\end{equation}
\begin{equation}
\Delta\phi_{Pu\pm} = \arccos\Bigg(  \sqrt{\frac{-3 \nu_\mathrm{v} \tau_\mathrm{d} + 6 \nu_\mathrm{v}  + 6 \tau_\mathrm{d} (\gamma_\mathrm{a} \nu_\mathrm{p} + \nu_\mathrm{d}) \mp A}{18\gamma_\mathrm{a} \nu_\mathrm{p} \nu_\mathrm{d}  \tau_\mathrm{d}^2/(\nu_\mathrm{d} \tau_\mathrm{d} + \nu_\mathrm{v})}} \Bigg)\, ,
\label{eq:omega0pm}
\end{equation}
where we introduced the abbreviation
\begin{equation}
\begin{aligned}
A = \Big[ &36\gamma_\mathrm{a}^2\nu_\mathrm{p}^2 \tau_\mathrm{d}^2  - 72 \gamma_\mathrm{a} \nu_\mathrm{p} \nu_\mathrm{d} \tau_\mathrm{d}^2  - 36 \gamma_\mathrm{a} \nu_\mathrm{p} \nu_\mathrm{v} \tau_\mathrm{d}^2  - 72 \gamma_\mathrm{a} \nu_\mathrm{p} \nu_\mathrm{v} \tau_\mathrm{d}  + 36 \nu_\mathrm{d}^2 \tau_\mathrm{d}^2   \\
&- 36 \nu_\mathrm{d} \nu_\mathrm{v} \tau_\mathrm{d}^2  + 72 \nu_\mathrm{d} \nu_\mathrm{v} \tau_\mathrm{d}  + 9 \nu_\mathrm{v}^2 \tau_\mathrm{d}^2 - 36 \nu_\mathrm{v}^2 \tau_\mathrm{d}  + 36 \nu_\mathrm{v}^2 \Big]^{\frac{1}{2}}\, .
\end{aligned}
\end{equation}

Using these expressions, the associated overall flow velocity $\mathbf{v}$ is then obtained from Eq.~(\ref{eq:StokesEquationRescaled}) under homogeneous conditions. It performs persistent rotations as well,
\begin{equation}
\mathbf{v} = v_{0\pm} \begin{pmatrix} \cos(\omega_{0\pm} - \Delta\phi_{Pv\pm}) \\ \sin(\omega_{0\pm} - \Delta\phi_{Pv\pm}) \end{pmatrix}\, .
\end{equation}
Its amplitude $v_0$ and phase shift $\Delta\phi_{Pv\pm}$ are obtained from $P_{0\pm}$, $u_{0\pm}$, and $\Delta\phi_{Pu\pm}$ via
\begin{align}
v_{0\pm} &= \sqrt{\frac{\nu_\mathrm{p}^2}{\nu_\mathrm{v}^2}P_{0\pm}^2 + \frac{\nu_\mathrm{d}^2}{\nu_\mathrm{v}^2}u_{0\pm}^2 - 2\frac{\nu_\mathrm{d} \nu_\mathrm{p}}{\nu_\mathrm{v}^2}P_{0\pm} u_{0\pm} \cos(\Delta\phi_{Pu\pm})}\, , \, \\
\Delta\phi_{Pv} &= \arctan\bigg(\frac{\nu_\mathrm{d} u_{0\pm} \sin(\Delta\phi_{Pu\pm})}{\nu_\mathrm{p} P_{0\pm} - \nu_\mathrm{d} u_{0\pm} \cos(\Delta\phi_{Pu\pm})} \bigg)\, .
\label{eq:v0pm}
\end{align}

\end{widetext}

To summarize, these solutions represent rotating spatially uniform translational polar orientational order, overall flows, and displacements. The magnitudes of polar orientational order, velocity, and elastic memory fields are constant. Yet, their directions rotate with locked phase shifts, following each other. %
The sense of rotation results from spontaneous symmetry breaking.

As an example, Fig.~\ref{fig:RotSol}(a) shows snapshots of the corresponding dynamic scenario for a specific set of parameter values. We illustrate the situation on the case of solid-like systems ($\tau_\mathrm{d}\rightarrow\infty$), where corresponding analytical expressions simplify and are listed in Appendix~\ref{app:OscillatorySolution}. The rotation of the three vector fields together with fixed phase-shifts are clearly visible. Our results were confirmed by numerical solutions for smaller spatially periodic systems, see Appendix~\ref{app:NumericalMethods}. The simulations confirm the analytical calculation. In Fig.~\ref{fig:RotSol}(b), we display the evolution of the corresponding $x$-components of $\mathbf{P}$, $\mathbf{v}$, and $\mathbf{u}$.
In agreement with the overall rotation, these components show harmonic oscillations.
The $y$-components perform the same harmonic oscillations as the $x$-components, yet with a phase shift of $\pi/2$, see Eqs.~(\ref{eq:PolarForm}).

\section{Transitions between the different analytical solutions}

\begin{figure}
\includegraphics[width=0.999\linewidth]{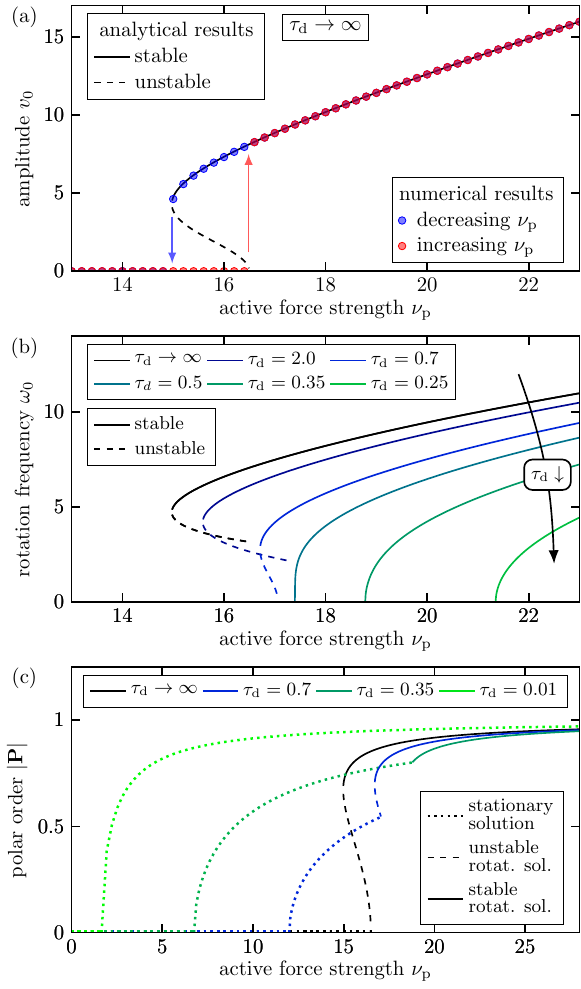}
\caption{\label{fig:RotSolViscoelastic}
Analysis of the transition between the different dynamic states. (a) Amplitude $v_0$ of the velocity, see Eqs.~(\ref{eq:PolarForm}), as a function of the strength $\nu_\mathrm{p}$ of active forcing for elastic solid-like systems ($\tau_\mathrm{d}\rightarrow\infty$). Analytical calculations (black lines) reveal a subcritical transition from the isotropic to the rotating state of polar orientational order. Here, solid lines denote stable and dashed lines unstable solutions. Numerical results (red and blue data points) confirm our analytical results and the associated hysteretic behavior. (b) Rotational frequency $\omega_0$ of the direction of spatially uniform motion as a function of $\nu_\mathrm{p}$ for the rotating solution, see Eqs.~(\ref{eq:PolarForm}). Results are displayed for decreasing relaxation time $\tau_\mathrm{d}$ of the memory of elastic displacements. This corresponds to declining elastic and increasing fluid-like behavior. Along this trend, we observe that the transition from quiescent to rotational dynamic states turns from sub- to supercritical. (c) Magnitude of polar orientational order $|\mathbf{P}|$ as a function of activity $\nu_\mathrm{p}$ for various values of the relaxation time $\tau_\mathrm{d}$ of the elastic memory. Isotropic and polar stationary solutions, see Eqs.~(\ref{eq:PolarSolution}), as well as stable and unstable rotational solutions are shown. The remaining parameters in all cases are set to $\gamma_\mathrm{a} = 1$, $\kappa = 0$, $\eta = 1$, $\mu = 1$, $\nu_\mathrm{v} = 1$, and $\nu_\mathrm{d} = 10$.}
\end{figure}

Mainly, we focus on the nature of the transitions between the different dynamic states. 
In fact, the rotating polar solution finally emerges when increasing the strength $\nu_\mathrm{p}$ of active forcing from the trivial isotropic solution of vanishing polar orientational order, velocity, and elastic memory field. 
To shed some light on this transition, %
we first consider the example of solid-like materials for $\tau_\mathrm{d} \to \infty$.
The remaining parameters are set to $\gamma_\mathrm{a} = 1$, $\kappa = 0$, $\eta = 1$, $\mu = 1$, $\nu_\mathrm{v} = 1$, and $\nu_\mathrm{d} = 10$.
We find that the rotational polar solution emerges from the isotropic state in a subcritical bifurcation when varying $\nu_\mathrm{p}$, see Fig.~\ref{fig:RotSolViscoelastic}(a).
There, we plot the magnitude $v_0$ of the velocity as a function of $\nu_\mathrm{p}$.

In fact, it turns out that the two solutions in Eqs.~(\ref{eq:P0pm})--(\ref{eq:omega0pm}) and (\ref{eq:v0pm}), distinguished by the sign of $\pm A$, correspond to the stable and unstable analytical solutions associated with this subcritical bifurcation. 
In Fig.~\ref{fig:RotSolViscoelastic}(a), stable and unstable branches are marked by solid and dashed lines, respectively. The presence of three branches of solution, one corresponding to the isotropic solution, one to the stable rotating polar solution, and one to the unstable rotating polar solution, indicates hysteretic behavior. 

We further investigate the subcritical transition to this rotational state by numerical calculations.
To this end, we solve Eqs.~(\ref{eq:EvolutionEquationPolarOrderRescaled})--%
(\ref{eq:EvolutionEquationDisplacementRescaled}) for small spatially periodic systems, %
see Appendix~\ref{app:NumericalMethods} for details.
The results are displayed in Fig.~\ref{fig:RotSolViscoelastic}(a) as the red and blue data points.
In the first set of calculations, we start from an isotropic, macroscopically quiescent state. Increasing the active forcing $\nu_\mathrm{p}$ (red dots), the system at a certain threshold jumps to the rotational solution of finite amplitudes $P_0$, $v_0$, and $u_0$.
However, when starting from an already rotating state and decreasing $\nu_\mathrm{p}$ (blue dots), the rotations persist down to a lower value of $\nu_\mathrm{p}$. The rotational solution vanishes at finite but smaller amplitudes $P_0$, $v_0$, and $u_0$.
This confirms the analytically identified region of hysteresis in a subcritical bifurcation scenario. Either a macroscopically quiescent, isotropic state or a solution of continuously rotating uniform spatial translation develops in this region, depending on the initial conditions. %

Next, we generalize our results from purely solid-like to viscoelastic behavior that allows for terminal flow. That is, finite values of the relaxation time $\tau_\mathrm{d}$ are considered. 
For the spatially uniform rotational solution, we find that the rotation frequency $\omega_0$ of the polar orientational order parameter, the velocity, and the elastic memory field depend on the strength $\nu_\mathrm{p}$ of active forcing and the relaxation time $\tau_\mathrm{d}$ of the memorized displacements. Examples for analytical solutions according to Eqs.~(\ref{eq:P0pm})--(\ref{eq:omega0pm}) are depicted in Fig.~\ref{fig:RotSolViscoelastic}(b). 
We recall that decreasing $\tau_\mathrm{d}$ moves the system from rather elastic solid-like towards more fluid-like behavior.

For all values of $\tau_\mathrm{d}$ shown in Fig.~\ref{fig:RotSolViscoelastic}(b), increasing the strength $\nu_\mathrm{p}$ of active forcing leads to faster rotations. 
Decreasing $\tau_\mathrm{d}$ (rendering the system less elastic), %
we first only observe quantitative changes. We focus on the point at which the rotating solution vanishes for decreasing $\nu_\mathrm{p}$. In Fig.~\ref{fig:RotSolViscoelastic}(b), this point is indicated by changing from the solid to the dashed line. The magnitude of the rotation frequency $\omega_0$ at this point becomes smaller for decreasing $\tau_\mathrm{d}$ (less elastic and more fluid-like behavior). Simultaneously, this point shifts to larger $\nu_\mathrm{p}$.
However, at $\tau_\mathrm{d} \approx 1$, the isotropic solution closely below the bifurcation is replaced by the nonrotating, stationary polar solution.

To illustrate this change, we show the magnitude of the analytically obtained polar order parameter $|\mathbf{P}|$ as a function of activity $\nu_\mathrm{p}$ for different relaxation times $\tau_\mathrm{d}$ of the elastic memory in Fig.~\ref{fig:RotSolViscoelastic}(c).
There, both stationary solutions (isotropic and polar) as well as the rotational solution are distinguished.
For example, for $\tau_\mathrm{d} = 0.7$, increasing $\nu_\mathrm{p}$ from zero first leads to a transition from the isotropic solution to a stationary state of global polar ordering and migration.
Then, rotations emerge in a subcritical transition. %
Remarkably, the transition from the stationary to the rotating state is accompanied by an increase in polar order. %

The situation again changes qualitatively at $\tau_\mathrm{d} \approx 0.5$.
There, the bifurcation changes from sub- to supercritical, see Figs.~\ref{fig:RotSolViscoelastic}(b) and (c).
Thus, for viscoelasticity of sufficiently low elastic contributions, that is, in a regime for small enough $\tau_\mathrm{d}$, the system can be driven continuously from the isotropic quiescent to the stationary polar and finally to the rotating polar dynamic state. 
There is no hysteresis for smaller values of $\tau_\mathrm{d}$.
Decreasing $\tau_\mathrm{d}$ even further, the rotating polar solution vanishes altogether, as is shown for $\tau_\mathrm{d} = 0.01$ in Fig.~\ref{fig:RotSolViscoelastic}(c).

Comparing the degree of polar orientational order for the stationary and for the rotational solution shown in Fig.~\ref{fig:RotSolViscoelastic}(c), we observe very different dependencies on the relaxation time $\tau_\mathrm{d}$.
In the stationary state, decreasing $\tau_\mathrm{d}$, and thus reducing elastic effects, generally leads to an increase in polar order.
Contrarily, the rotating state exhibits a higher degree of polar order with increasing $\tau_\mathrm{d}$.
However, for larger activities in the rotating state, this effect diminishes and the curves in Fig.~\ref{fig:RotSolViscoelastic}(c) approach each other.

Physically, sufficiently long relaxation times $\tau_\mathrm{d}$, that is, more pronounced elastic effects hinder the flow $\mathbf{v}$. The elastic restoring forces are too large in this case to allow for persistent migration in the same direction. As a compromise, rotations occur. Reducing the role of elasticity through decreasing $\tau_\mathrm{d}$ facilitates unbounded flows $\mathbf{v}$. Then, instead of rotations, persistent stationary flows combined with polar orientational order are observed.

\begin{figure}
\includegraphics[width=0.999\linewidth]{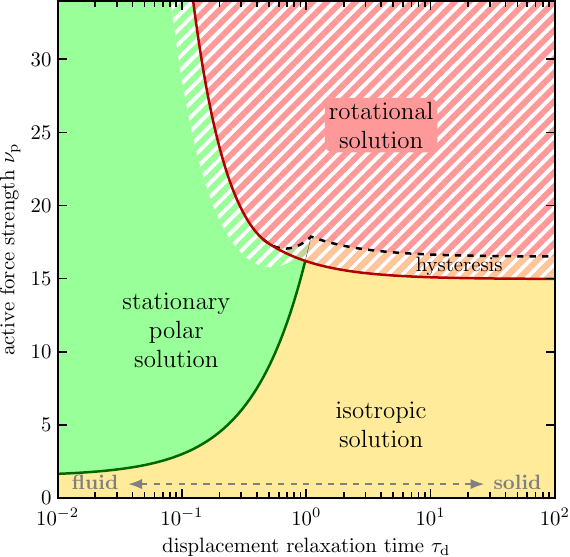}
\caption{\label{fig:StateDiagram}State diagram for small, idealized systems that combine activity and viscoelasticity. We here impose spatial uniformity of the solutions to enable analytical considerations for illustration of the theory. Dropping this idealization leads to a reconsideration of the state diagram and novel dynamic states as described in a companion paper \cite{reinken2025rheologically}. We hatch the areas of significant deviation when the idealization of spatial uniformity is dropped. Such deviations mainly occur when both activity and (visco)elastic effects become substantial, which corresponds to the major focus of our considerations. The diagram is displayed as a function of the relaxation time of elastic memory $\tau_\mathrm{d}$, which measures the importance of elastic effects, and the strength of active forcing $\nu_\mathrm{p}$. Green and yellow areas denote the regions of uniform stationary polar and isotropic solutions, respectively. The spatially homogeneous rotational solution exists above the red curve.
A hysteretic region is indicated by the dashed black line. We set the remaining parameters to $\gamma_\mathrm{a} = 1$, $\kappa = 0$, $\eta = 1$, $\mu = 1$, $\nu_\mathrm{v} = 1$, and $\nu_\mathrm{d} = 10$.}
\end{figure}

The boundaries between the three simplified uniform solutions are further visualized in Fig.~\ref{fig:StateDiagram}, which shows a state diagram in the plane spanned by the elastic relaxation time $\tau_\mathrm{d}$ and the strength of active forcing $\nu_\mathrm{p}$.
To summarize, the stationary polar solution is encountered for fluid-like systems of smaller $\tau_\mathrm{d}$ at elevated activity, while the disordered isotropic solution is found at lower activities.
The rotational solution, see Eqs.~(\ref{eq:P0pm})--(\ref{eq:v0pm}), emerges for more solid-like systems, that is, at larger $\tau_\mathrm{d}$, and at higher activity.
Between the rotational solution and the two stationary states, as discussed above and in Fig.~\ref{fig:RotSolViscoelastic}, a hysteretical regime is observed at the lower boundary of this area.
Our analytical results and the state diagram shown in Fig.~\ref{fig:StateDiagram} refer to idealized systems of small size under periodic boundary conditions, where we enforce the solutions to remain spatially uniform. 
The more general case is discussed in a companion article~\cite{reinken2025rheologically}, where we demonstrate that complex spatiotemporal patterns emerge in spatially extended systems. Deviations occur specifically in the regime of more pronounced activity and (visco)elastic effects, which forms our main subject in this work. We mark these areas of complex novel dynamic states by hatches in Fig.~\ref{fig:StateDiagram}.

\section{Conclusions}
\label{sec:Conclusions}

Our goal was to derive and explore a theoretical description for active suspensions that is continuously tunable from viscous fluid- to elastic solid-like behavior, with a broad spectrum of viscoelastic properties in between.
Using analytical calculations, we identify three different spatially uniform solutions.
At low activities, the system does not display any global polar orientational order and remains isotropic.
Increasing activity, different dynamic states emerge, depending on whether the system is more fluid- or more solid-like.

In the more fluid-like case, we find a spatially uniform state of global polar orientational order, similar to what is observed in conventional theories for nonelastic systems~\cite{vicsek1995novel,toner1998flocks,toner2005hydrodynamics}.
In contrast to that, solid-like properties prevent persistent global motion in one direction due to the associated elastic restoring forces.
There, we observe a spatially uniform state of rotating global polar orientational order. It represents globally uniform motion in one direction, while this direction rotates.
When increasing the activity starting from the disordered, isotropic state, these rotations emerge in a subcritical transition exhibiting hysteresis.
Our unified description reveals that this hysteretic regime vanishes as the bifurcation changes from sub- to supercritical when decreasing the relaxation time of the memory of elastic displacements $\tau_\mathrm{d}$, that is, turning to more fluid-like behavior in the viscoelastic regime.
Remarkably, with increasing activity, the dynamic state can transition from isotropic to stationary polar and then to rotating polar order.

Our description bridges the gap between active fluids and active solids, which have by now mostly been considered in separate frameworks. 
As pointed out in Ref.~\onlinecite{nijjer2023bacteria}, there is a critical need for this kind of integrative theories particularly for biological media, as they may change their properties dynamically and continuously. 
For example, during biofilm formation, bacteria excrete extracellular polymers~\cite{hall2004bacterial,worlitzer2022biophysical,jana2020nonlinear}. Accordingly, the film starts from still fluid-like, but over time changes its rheological properties from viscous and viscoelastic to increasingly elastic.
In a settled biofilm, the polymer network forms an elastic matrix and the resulting system may then be described as an active solid~\cite{xu2023autonomous}.
On a continuum-theoretical level, our description covers such transitions between fluid- and solid-like behavior. It shall be applied in future studies accordingly.

Still, our description represents only a first step towards a generalized continuum-theoretical framework of active viscoelastic media.
We here focus on the effects of active driving forces for migrating active units embedded in the viscoelastic medium near a substrate.
Next, taking active stresses into account is straightforward~\cite{simha2002hydrodynamic,saintillan2008instabilities,reinken2018derivation} and a promising route for future research.
For example, the effects of viscoelasticity on mesoscale turbulence~\cite{wensink2012meso,alert2021active,reinken2024pattern} shall be explored in this way, under varying conditions of viscoelasticity.

In view of possible additional extensions of our investigations, we mention the alignment of the active units with the flow field. Here, we considered the situation in which the self-propelled objects tend to align along the direction of surrounding flow $\mathbf{v}$, ($\gamma_\mathrm{a}>0$), because it appears most natural.
The opposite case ($\gamma_\mathrm{a}<0$)
benefits antialignment. Possibly, this situation may enhance further dynamical states, also in view of 
flocking and clustering transitions and active transport~\cite{zhang2021active,das2024flocking,casiulis2024geometric,arbel2024mechanical} when we leave our assumption of low to moderate densities. 

We remark that, naturally, the rotating state reminds a bit of the dynamics of chiral, self-rotating active particles~\cite{liebchen2022chiral}. However, at second glance, the underlying framework is very different. Chiral active particles by themselves tend to persistently bend their trajectories and rotate their body while migrating. Often, the necessary underlying symmetry breaking is permanent, leading to a predefinded sense of rotation.
In contrast to that, rotation in our case results from spontaneous symmetry breaking. Particularly, it is an immediate consequence of the coupling between activity and (visco)elasticity. Our self-propelled objects, without perturbations, would migrate along straight trajectories. Circling is induced by the elastic restoring forces. They hinder persistent motion in one direction. 
As a compromise, localized circling around a fixed spot occurs. 
Still, chiral active particles show various remarkable phenomena~\cite{liebchen2022chiral}, for example, the formation of disordered hyperuniform structures~\cite{lei2019nonequilibrium}.
Future research may additionally investigate the behavior of chiral active units embedded in viscoelastic background media using our approach.

Further extensions concern 
higher densities. Then, mutual steric and other short-ranged interactions become more significant. Corresponding situations require suitable statistical considerations for developing the theoretical description by coarse-graining from the discrete particle level. Examples are dynamical density functional theories for active objects~\cite{menzel2014active, hoell2019multi}, Boltzmann-equation-type constructions~\cite{bertin2006boltzmann,bertin2009hydrodynamic} and related approaches~\cite{dean1996langevin}. Their coupling to the generalized hydrodynamic theory of viscoelasticity \cite{temmen2000convective, pleiner2004nonlinear, menzel2025linear} is an obvious task for the future.

\begin{acknowledgments}
We thank the Deutsche Forschungsgemeinschaft (German Research Foundation, DFG) for support through the Research Grant no.\ ME 3571/5-1. Moreover, A.M.M. acknowledges support through the Heisenberg Grant no.\ ME 3571/4-1. 
\end{acknowledgments}

\section*{Data Availability Statement}
The code used to generate the data in support of the findings of this article is openly available~\cite{code2025unified}. 

\appendix

\section{Elastic stress and memory field of elastic displacements}
\label{app:DisplacementFieldDynamics}

To ensure clarity for all readers, particularly those who mainly work with descriptions of the Lagrange type and have not been in contact with the Euler perspective for a while, we recall some of the basic items of the well-established background \cite{martin1972unified, temmen2000convective} of Eq.~(\ref{eq:EvolutionEquationDisplacement}). Generally, Lagrange descriptions are trajectory-based. We fix our attention to the material elements and observe how they move. The trajectory of a material element, $\mathbf{R}(t)$, is obtained from the Euler frame by integrating over time the velocity field at its position at each moment in time, $\mathbf{R}(t)=\int_{t_0}^t\mathbf{v}\left(\mathbf{R}(t'),t'\right)\mathrm{d}t'$. $t_0$ marks the time of the beginning of our consideration. Thus, we assign to any material element located at time $t_0$ at position $\mathbf{R}(t_0)$ where we will find it later, namely, $\mathbf{R}(t)$. The distance between the two positions can be considered as displacement.

In the Euler framework, the perspective is different. We now describe the system in its the current, possibly deformed state at time $t$, with the frame of reference fixed to the physical space (not the material). 
Thus, for any position in space $\mathbf{x}$ that we associate with the deformed state of an elastic material, we describe from which position $\mathbf{x}-\mathbf{u}(\mathbf{x},t)$ the material element currently located at $\mathbf{x}$ at time $t$ has come from.
Therefore, the field $\mathbf{u}$ can be considered as an elastic memory of the previous positions. The material elements tend to displace back to these positions, if external or internal forces causing the deformation were released.

If the system is not perfectly elastic, induced elastic stresses or strains resulting from deformations may relax over time, although the overall shape of the system is maintained in the deformed state. This may happen through internal restructuring or structural rearrangement. An illustrative example are entangled polymer melts that can relax their internal elastic stresses and strains through disentanglement processes \cite{strobl1997physics} (we recall that, particularly for linear elasticity and in generalized hydrodynamics \cite{temmen2000convective}, elastic stresses and strains in the Euler framework are equivalent due to elastic stress-strain relations). This relaxation process is one manifestation of viscoelasticity. It means that the positions to which the material elements tend to displace back to have approached the current space positions. The former are given by $\mathbf{x}-\mathbf{u}(\mathbf{x},t)$, the latter by $\mathbf{x}$. Thus, when the former approach the latter, $\mathbf{u}$ decays over time. The decay rate is called $\tau_\mathrm{d}^{-1}$ in Eq.~(\ref{eq:EvolutionEquationDisplacement}). We have described these relations in an illustrative way here. Mathematically, they follow in a straightforward way from the systematic generalized hydrodynamics approach \cite{temmen2000convective, pleiner2004nonlinear, puljiz2019memory, menzel2025linear}.

For more convenient reference, we summarize the background of Eq.~(\ref{eq:EvolutionEquationDisplacement}). The corresponding systematic generalized hydrodynamics theory was derived and presented about a quarter of a century ago \cite{temmen2000convective, pleiner2004nonlinear}.
We start with the static part of the generalized hydrodynamics theory, where we here, for brevity, confine ourselves to relations associated with (visco)elasticity. This concerns the elastic stress-strain relation. It follows from the elastic energy density $\epsilon$ in the Eulerian frame of reference, 
which reads
\begin{equation}
\epsilon = \frac{1}{2}K_{ijkm} {U}^\mathrm{nonlin}_{ij} {U}^\mathrm{nonlin}_{km}\, .
\end{equation}
Here, the components of the nonlinear strain tensor in terms of the memory field $\mathbf{u}$ in the Eulerian frame \cite{chaikin1995principles, temmen2000convective} are given by
\begin{equation}
\label{eq:DeformationGradient}
{U}^\mathrm{nonlin}_{ij} = \frac{1}{2}\big[ \partial_{x_i} u_j + \partial_{x_j} u_i - (\partial_{x_i} u_k) (\partial_{x_j} u_k)  \big]\, .
\end{equation}
We remark that the minus sign becomes a plus sign in the Lagrange frame \cite{chaikin1995principles}.
The fourth-rank tensor $\mathbf{K}$ is the stiffness tensor, which for an isotropic material is of the form
\begin{equation}
K_{ijkm} = \mu (\delta_{ik}\delta_{jm}  + \delta_{im}\delta_{jk} )+\lambda\delta_{ij}\delta_{km}\, .
\end{equation}
$\delta_{ij}$ denotes the Kronecker delta, $\mu$ is the shear modulus, and the material parameter $\lambda$ diverges for incompressible materials. 
We here confine ourselves to orders linear in strain, that is, linear in gradients $\nabla\mathbf{u}$. Thus, the components of the strain tensor reduce to  
\begin{equation}\label{eq:strainlin}
U_{ij} = \frac{1}{2}\big[ \partial_{x_i} u_j + \partial_{x_j} u_i  \big]\, .
\end{equation}

In our considerations, we assume the materials to be incompressible. 
The diverging material parameter $\lambda$ leads to a vanishing trace of $\mathbf{U}$ to linear order, that is, $\nabla\cdot\mathbf{u}=0$. Ensuring that this condition remains satisfied, see below, allows us to drop the contribution of $\lambda$ in the following. 

An infinitesimal change of $\epsilon$ with $\mathbf{U}$, 
\begin{equation}
d \epsilon = \Psi_{ij} d U_{ij}\,  ,
\end{equation}
allows us to write the components of the associated thermodynamic conjugate $\boldsymbol{\Psi}$ as
\begin{equation}
\Psi_{ij} = \mu (U_{ij} + U_{ji}) = 2\mu U_{ij}\, ,
\end{equation}
where the latter equality applies because $\mathbf{U}$ is symmetric. 

The elastic shear forces enter the flow equation for the velocity field $\mathbf{v}$, Eq.~(\ref{eq:StokesEquation}), through the divergence of the elastic stress tensor $\boldsymbol{\sigma}^\mathrm{el}$, see also the paragraph below Eq.~(\ref{eq:StokesEquation}).  The components of $\boldsymbol{\sigma}^\mathrm{el}$ read \cite{temmen2000convective} 
\begin{equation}
\label{eq:elasticStress}
{\sigma}^\mathrm{el}_{ij} = \Psi_{ij} - \Psi_{ki} U_{jk} - \Psi_{kj} U_{ik}\, .
\end{equation}
We here confine ourselves to terms linear in elastic strain, so that
\begin{equation}
\label{eq:elasticStressLinear}
{\sigma}^\mathrm{el}_{ij} = \Psi_{ij} = 2\mu U_{ij}\, . %
\end{equation}
The linearized components of the elastic stress tensor $\sigma_{ij}^\mathrm{el}$ are thus obtained as 
\begin{equation}
\sigma_{ij}^\mathrm{el} = \mu (\partial_{x_i} u_j + \partial_{x_j} u_i)\, 
\end{equation}
in terms of the memory field of elastic displacements $\mathbf{u}$. Due to incompressibility, $\nabla\cdot\bm{\sigma}^\mathrm{el}=\mu\nabla^2\mathbf{u}$, which enters the flow equation for $\mathbf{v}$, Eq.~(\ref{eq:StokesEquation}).

\begin{widetext}

Second, the dynamic part of the generalized hydrodynamics theory associated with (visco)elasticity \cite{temmen2000convective, pleiner2004nonlinear} is the dynamic equation for the strain tensor. In its vectorial notation, that is, for the elastic memory field $\mathbf{u}$, it corresponds to our Eq.~(\ref{eq:EvolutionEquationDisplacement}). To begin, this dynamic equation in Ref.~\onlinecite{temmen2000convective} linearized in strain $\mathbf{U}$ reads
\begin{equation}
\label{eq:EvolutionEquationDeformationGradient}
\begin{aligned}
\partial_t \mathbf{U} + \mathbf{v} \cdot \nabla \mathbf{U} + (\nabla \mathbf{v}) \cdot \mathbf{U} + \mathbf{U} \cdot (\nabla \mathbf{v})^\top = \boldsymbol{\Sigma} - \tau_\mathrm{d}^{-1} \mathbf{U}\, .
\end{aligned}
\end{equation}
It includes as the second term on the left-hand side the translational convective transport with the flow field $\mathbf{v}$. %
Lower convective transport is described by the subsequent two terms \cite{pleiner2004nonlinear}, where $^\top$ marks the transpose. On the right-hand side, strains are generated by flows, that is, motion of material, according to the deformation rate tensor $\boldsymbol{\Sigma} = [(\nabla \mathbf{v})^\top + (\nabla \mathbf{v})]/2$. 
The form of the equation can be rigorously derived from conservation laws and symmetry arguments~\cite{temmen2000convective}. Without the relaxation term ${}- \tau_\mathrm{d}^{-1} \mathbf{U}$, Eq.~(\ref{eq:EvolutionEquationDeformationGradient}) describes genuinely elastic systems. We again assume isotropic elasticity as well as incompressibility, that is, $U_{ii} = 0$. An additional relaxation term, here denoted as ${}- \tau_\mathrm{d}^{-1} \mathbf{U}$, enters the right-hand side in the case of viscoelastic systems that are not strictly elastic solid-like~\cite{temmen2000convective}. In our framework, it describes the loss of memory about previous undeformed states over time. Thus, when elastic stresses are released, the material elements are not fully displaced back to their initial positions, but to some intermediate locations. In other words, the material has flown a certain distance. 
$\tau_\mathrm{d}$ denotes the corresponding relaxation time.

We now insert into Eq.~(\ref{eq:EvolutionEquationDeformationGradient}) the expression for the linearized strain tensor, Eq.~(\ref{eq:strainlin}).
Accordingly, the field $\mathbf{u}$ 
is interpreted in an Eulerian framework as well.
Thus, it describes the memory of the reversible elastic displacements that are presently stored in the material for the material elements currently located at positions $\mathbf{x}$ at time $t$.
In other words, if all imposed stresses were switched off at time $t$, the material element currently located at position $\mathbf{x}$ at time $t$ tends to move back to $\mathbf{x} - \mathbf{u}(\mathbf{x},t)$.
From Eq.~(\ref{eq:EvolutionEquationDeformationGradient}), we obtain
\begin{equation}
\label{eq:EvolutionEquationDisplacement1}
\begin{aligned}
\partial_t \partial_{x_i} u_j &+ \partial_t \partial_{x_j} u_i +
v_k \partial_{x_k} \partial_{x_i} u_j + v_k \partial_{x_k} \partial_{x_j} u_i + (\partial_{x_i} v_k) (\partial_{x_k} u_j) + (\partial_{x_i} v_k) (\partial_{x_j} u_k ) \\ &+ (\partial_{x_j} v_k) (\partial_{x_i} u_k) + (\partial_{x_j} v_k) (\partial_{x_k} u_i ) 
= \partial_{x_i} v_j + \partial_{x_j} v_i - {\tau_\mathrm{d}}^{-1}(\partial_{x_i} u_j + \partial_{x_j} u_i)\, .
\end{aligned}
\end{equation}
We aim for a consistent theory in the incompressible limit, that is, $\nabla\cdot\mathbf{v}=\partial_{x_i} v_i = 0$ and $\nabla\cdot\mathbf{u}=\partial_{x_i} u_i = 0$ must be satisfied at all times.
Taking the trace of Eq.~(\ref{eq:EvolutionEquationDisplacement1}) yields an evolution equation for $\nabla\cdot\mathbf{u}$, 
\begin{equation}
\label{eq:EvolutionEquationDisplacementTrace}
\begin{aligned}
\partial_t \partial_{x_i} u_i +
v_k \partial_{x_k} \partial_{x_i} u_i + (\partial_{x_i} v_k) (\partial_{x_k} u_i) + (\partial_{x_i} v_k) (\partial_{x_i} u_k ) 
= - {\tau_\mathrm{d}}^{-1}\partial_{x_i} u_i\, ,
\end{aligned}
\end{equation}
where we already used $\nabla\cdot\mathbf{v}=\partial_{x_i} v_i = 0$.
We find that only the nonlinear terms $(\partial_{x_i} v_k) (\partial_{x_k} u_i)$ and $(\partial_{x_i} v_k) (\partial_{x_i} u_k ) $ in Eq.~(\ref{eq:EvolutionEquationDisplacementTrace}) associated with rotational convection may violate $\partial_{x_i} u_i = 0$ and thus may create a nonvanishing $\nabla\cdot\mathbf{u}$. %
These considerations imply that the nonlinear coupling terms between $\nabla\mathbf{v}$ and $\nabla\mathbf{u}$ associated with rotational convection are not in line with the linearized description in $\nabla\mathbf{u}$. 
We must therefore drop these higher-order terms in our equations, leading to
\begin{equation}
\label{eq:EvolutionEquationDisplacement2}
\partial_t \partial_{x_i} u_j + \partial_t \partial_{x_j} u_i +
v_k \partial_{x_k} \partial_{x_i} u_j + v_k \partial_{x_k} \partial_{x_j} u_i = \partial_{x_i} v_j + \partial_{x_j} v_i - {\tau_\mathrm{d}}^{-1}(\partial_{x_i} u_j + \partial_{x_j} u_i)\, .
\end{equation}
Next, we express the convective terms via
\begin{equation}
\label{eq:EvolutionEquationDisplacementAdvection}
\begin{aligned}
v_k \partial_{x_k} \partial_{x_i} u_j &= \partial_{x_i} (v_k \partial_{x_k} u_j) - (\partial_{x_i} v_k)(\partial_{x_k} u_j)\, , \\ 
v_k \partial_{x_k} \partial_{x_j} u_i &= \partial_{x_j} (v_k \partial_{x_k} u_i) - (\partial_{x_j} v_k)(\partial_{x_k} u_i)\, .
\end{aligned}
\end{equation}
Inserting into Eq.~(\ref{eq:EvolutionEquationDisplacement1}) yields
\begin{equation}
\label{eq:EvolutionEquationDisplacement5}
\begin{aligned}
\partial_{x_i} \big[ \partial_t u_j +  v_k \partial_{x_k} u_j - v_j + \tau_\mathrm{d}^{-1} u_j \big] + \partial_{x_j} \big[ \partial_t u_i +  v_k \partial_{x_k} u_i - v_i + \tau_\mathrm{d}^{-1} u_i \big]\\ 
= (\partial_{x_i} v_k)(\partial_{x_k} u_j) + (\partial_{x_j} v_k)(\partial_{x_k} u_i)  \, .
\end{aligned}
\end{equation}
To deal with the second line of Eq.~(\ref{eq:EvolutionEquationDisplacement5}), we introduce the auxiliary variable $q$ by adding $2 \partial_{x_i}\partial_{x_i} q$ on both sides,
\begin{equation}
\begin{aligned}
\partial_{x_i} \big[ \partial_t u_j +  v_k \partial_{x_k} u_j - v_j + \tau_\mathrm{d}^{-1} u_j + \partial_{x_j} q \big] + \partial_{x_j} \big[ \partial_t u_i +  v_k \partial_{x_k} u_i - v_i + \tau_\mathrm{d}^{-1} u_i + \partial_{x_i} q \big]\\ = (\partial_{x_i} v_k)(\partial_{x_k} u_j) + (\partial_{x_j} v_k)(\partial_{x_k} u_i) + 2 \partial_{x_i} \partial_{x_j} q \, .
\end{aligned}
\label{eq:lhs-rhs}
\end{equation}

\end{widetext}

For vanishing right-hand side, the left-hand side is consistent with Eq.~(\ref{eq:EvolutionEquationDisplacement}). Equation~(\ref{eq:EvolutionEquationDisplacement}) only contains couplings between the velocity itself (not its gradient) and gradients of the memory field of elastic displacements $\mathbf{u}$. 
On the right-hand side of Eq.~(\ref{eq:lhs-rhs}), we find nonlinear couplings between velocity gradients and gradients of the elastic memory field $\mathbf{u}$, as well as the auxiliary field $q$. Since we 
confine ourselves to linear elasticity, meaning linearity in gradients of $\mathbf{u}$, 
the description shall maintain our condition of incompressibility $\nabla\cdot\mathbf{u}=0$ within the linear regime. 
To this end, we take the trace of Eq.~(\ref{eq:lhs-rhs}) and enforce its right-hand side to vanish, which leads to 
an equation for the auxiliary field $q$,
\begin{equation}
\label{eq:AuxiliaryVariable}
\partial_{x_i}^2 q = - (\partial_{x_i} v_k)(\partial_{x_k} u_i) \, .
\end{equation}
It serves to determine the field $q$ and  together with Eq.~(\ref{eq:EvolutionEquationDisplacement}) %
describes the evolution of the elastic memory field $\mathbf{u}$ in a consistent way when confining ourselves to linear order in $\nabla\mathbf{u}$ and incompressible systems. 

Physically, these considerations imply that the nonlinear term associated with translational convection of the elastic memory field $\mathbf{u}$ is maintained. This is one step beyond ordinary linear elasticity.
In particular, this contribution becomes important when flows arise for viscoelastic fluids and therefore is physically significant.

\section{Spatially uniform polar rotational solutions for elastic systems}
\label{app:OscillatorySolution}

In the following, we summarize the expressions corresponding to Eqs.~(\ref{eq:P0pm})--(\ref{eq:omega0pm}) in the elastic, solid-like limit ($\tau_\mathrm{d} \to \infty$). These expressions simplify to
\begin{equation}
P_{0\pm} = \sqrt{\frac{-4\gamma_\mathrm{a} \nu_\mathrm{p}  + 8 \nu_\mathrm{d}  + 2 \nu_\mathrm{v} \pm 2B}{2 \gamma_\mathrm{a} \nu_\mathrm{p}  + 2 \nu_\mathrm{d}  - \nu_\mathrm{v} \mp B}}\, ,
\end{equation}
\begin{equation}
u_{0\pm} = \sqrt{\frac{\nu_\mathrm{p} (-2\gamma_\mathrm{a} \nu_\mathrm{p}  + 4 \nu_\mathrm{d}  + \nu_\mathrm{v} \pm B)}{2 \gamma_\mathrm{a} \nu_\mathrm{d}^2 }}\, ,
\end{equation}
\begin{equation}
\omega_{0\pm} = \frac{\nu_\mathrm{d}}{\nu_\mathrm{v}}\sqrt{\frac{2 \gamma_\mathrm{a} \nu_\mathrm{p}  - 2 \nu_\mathrm{d}  + \nu_\mathrm{v} \pm B}{2 \gamma_\mathrm{a} \nu_\mathrm{p}  + 2 \nu_\mathrm{d}  - \nu_\mathrm{v} \mp B}}\, ,
\end{equation}
\begin{equation}
\Delta\phi_{Pu\pm} = \arccos\Bigg( \sqrt{ \frac{2\gamma_\mathrm{a} \nu_\mathrm{p}  + 2 \nu_\mathrm{d}  - \nu_\mathrm{v} \mp B}{4 \gamma_\mathrm{a} \nu_\mathrm{p}  } } \Bigg)\, ,
\end{equation}
where we defined the abbreviation
\begin{equation}
B = \sqrt{4\gamma_\mathrm{a}^2 \nu_\mathrm{p}^2  - 8 \gamma_\mathrm{a} \nu_\mathrm{p} \nu_\mathrm{d}  - 4 \gamma_\mathrm{a} \nu_\mathrm{p} \nu_\mathrm{v}  + 4 \nu_\mathrm{d}^2  - 4 \nu_\mathrm{d} \nu_\mathrm{v}  + \nu_\mathrm{v}^2}\, .
\end{equation}

\section{Numerical methods}
\label{app:NumericalMethods}

We use a pseudo-spectral method~\cite{canuto2007spectral} to solve Eqs.~(\ref{eq:EvolutionEquationPolarOrderRescaled})--%
(\ref{eq:EvolutionEquationDisplacementRescaled}) in a two-dimensional domain under periodic boundary conditions.
This scheme exploits the fast Fourier transformation to calculate spatial derivatives, which significantly accelerates solving Eq.~(\ref{eq:StokesEquationRescaled}).
Time integration is performed via a fourth-order Runge--Kutta method using a time step of $\Delta t = 10^{-3}$ in our rescaled units.
To ensure that the velocity field $\mathbf{v}$ stays divergence-free, we use a projection method~\cite{durran2010numerical}. The pressure $p$ is determined in a way to satisfy the incompressibility conditions.
If not stated otherwise, our numerical calculations start from a weakly perturbed isotropic state. Random initial values taken from a uniform distribution over the interval $[-0.01,0.01]$ are assigned to each of the components of the polar orientational order parameter field at every grid point.
Velocity and elastic memory fields are set to zero initially.
When investigating hysteretic behavior, we successively change the activity as quantified by the parameter $\nu_\mathrm{p}$, see Eq.~(\ref{eq:AciveForcing}), and use the results of the previous calculation as a starting point for the next step. 
To obtain the data shown in Fig.~\ref{fig:RotSol} and Fig~\ref{fig:RotSolViscoelastic}(a), we use system sizes of $5 \times 5$ in rescaled units, spanned by a numerical grid of $32 \times 32$ points.
We analyze the dynamics in a window of $1000$ time units to measure each of the displayed data points in Fig~\ref{fig:RotSolViscoelastic}(a).


\begin{thebibliography}{89}%
\makeatletter
\providecommand \@ifxundefined [1]{%
 \@ifx{#1\undefined}
}%
\providecommand \@ifnum [1]{%
 \ifnum #1\expandafter \@firstoftwo
 \else \expandafter \@secondoftwo
 \fi
}%
\providecommand \@ifx [1]{%
 \ifx #1\expandafter \@firstoftwo
 \else \expandafter \@secondoftwo
 \fi
}%
\providecommand \natexlab [1]{#1}%
\providecommand \enquote  [1]{``#1''}%
\providecommand \bibnamefont  [1]{#1}%
\providecommand \bibfnamefont [1]{#1}%
\providecommand \citenamefont [1]{#1}%
\providecommand \href@noop [0]{\@secondoftwo}%
\providecommand \href [0]{\begingroup \@sanitize@url \@href}%
\providecommand \@href[1]{\@@startlink{#1}\@@href}%
\providecommand \@@href[1]{\endgroup#1\@@endlink}%
\providecommand \@sanitize@url [0]{\catcode `\\12\catcode `\$12\catcode
  `\&12\catcode `\#12\catcode `\^12\catcode `\_12\catcode `\%12\relax}%
\providecommand \@@startlink[1]{}%
\providecommand \@@endlink[0]{}%
\providecommand \url  [0]{\begingroup\@sanitize@url \@url }%
\providecommand \@url [1]{\endgroup\@href {#1}{\urlprefix }}%
\providecommand \urlprefix  [0]{URL }%
\providecommand \Eprint [0]{\href }%
\providecommand \doibase [0]{https://doi.org/}%
\providecommand \selectlanguage [0]{\@gobble}%
\providecommand \bibinfo  [0]{\@secondoftwo}%
\providecommand \bibfield  [0]{\@secondoftwo}%
\providecommand \translation [1]{[#1]}%
\providecommand \BibitemOpen [0]{}%
\providecommand \bibitemStop [0]{}%
\providecommand \bibitemNoStop [0]{.\EOS\space}%
\providecommand \EOS [0]{\spacefactor3000\relax}%
\providecommand \BibitemShut  [1]{\csname bibitem#1\endcsname}%
\let\auto@bib@innerbib\@empty
\bibitem [{\citenamefont {Marchetti}\ \emph {et~al.}(2013)\citenamefont
  {Marchetti}, \citenamefont {Joanny}, \citenamefont {Ramaswamy}, \citenamefont
  {Liverpool}, \citenamefont {Prost}, \citenamefont {Rao},\ and\ \citenamefont
  {Simha}}]{marchetti2013hydrodynamics}%
  \BibitemOpen
  \bibfield  {author} {\bibinfo {author} {\bibfnamefont {M.~C.}\ \bibnamefont
  {Marchetti}}, \bibinfo {author} {\bibfnamefont {J.~F.}\ \bibnamefont
  {Joanny}}, \bibinfo {author} {\bibfnamefont {S.}~\bibnamefont {Ramaswamy}},
  \bibinfo {author} {\bibfnamefont {T.~B.}\ \bibnamefont {Liverpool}}, \bibinfo
  {author} {\bibfnamefont {J.}~\bibnamefont {Prost}}, \bibinfo {author}
  {\bibfnamefont {M.}~\bibnamefont {Rao}},\ and\ \bibinfo {author}
  {\bibfnamefont {R.~A.}\ \bibnamefont {Simha}},\ }\bibfield  {title} {\bibinfo
  {title} {Hydrodynamics of soft active matter},\ }\href@noop {} {\bibfield
  {journal} {\bibinfo  {journal} {Rev. Mod. Phys.}\ }\textbf {\bibinfo {volume}
  {85}},\ \bibinfo {pages} {1143} (\bibinfo {year} {2013})}\BibitemShut
  {NoStop}%
\bibitem [{\citenamefont {Bechinger}\ \emph {et~al.}(2016)\citenamefont
  {Bechinger}, \citenamefont {Di~Leonardo}, \citenamefont {L\"owen},
  \citenamefont {Reichhardt}, \citenamefont {Volpe},\ and\ \citenamefont
  {Volpe}}]{bechinger2016}%
  \BibitemOpen
  \bibfield  {author} {\bibinfo {author} {\bibfnamefont {C.}~\bibnamefont
  {Bechinger}}, \bibinfo {author} {\bibfnamefont {R.}~\bibnamefont
  {Di~Leonardo}}, \bibinfo {author} {\bibfnamefont {H.}~\bibnamefont
  {L\"owen}}, \bibinfo {author} {\bibfnamefont {C.}~\bibnamefont {Reichhardt}},
  \bibinfo {author} {\bibfnamefont {G.}~\bibnamefont {Volpe}},\ and\ \bibinfo
  {author} {\bibfnamefont {G.}~\bibnamefont {Volpe}},\ }\bibfield  {title}
  {\bibinfo {title} {Active particles in complex and crowded environments},\
  }\href@noop {} {\bibfield  {journal} {\bibinfo  {journal} {Rev. Mod. Phys.}\
  }\textbf {\bibinfo {volume} {88}},\ \bibinfo {pages} {045006} (\bibinfo
  {year} {2016})}\BibitemShut {NoStop}%
\bibitem [{\citenamefont {Lauga}\ and\ \citenamefont
  {Powers}(2009)}]{lauga2009hydrodynamics}%
  \BibitemOpen
  \bibfield  {author} {\bibinfo {author} {\bibfnamefont {E.}~\bibnamefont
  {Lauga}}\ and\ \bibinfo {author} {\bibfnamefont {T.~R.}\ \bibnamefont
  {Powers}},\ }\bibfield  {title} {\bibinfo {title} {The hydrodynamics of
  swimming microorganisms},\ }\href@noop {} {\bibfield  {journal} {\bibinfo
  {journal} {Rep. Prog. Phys.}\ }\textbf {\bibinfo {volume} {72}},\ \bibinfo
  {pages} {096601} (\bibinfo {year} {2009})}\BibitemShut {NoStop}%
\bibitem [{\citenamefont {Elgeti}\ \emph {et~al.}(2015)\citenamefont {Elgeti},
  \citenamefont {Winkler},\ and\ \citenamefont {Gompper}}]{elgeti2015physics}%
  \BibitemOpen
  \bibfield  {author} {\bibinfo {author} {\bibfnamefont {J.}~\bibnamefont
  {Elgeti}}, \bibinfo {author} {\bibfnamefont {R.~G.}\ \bibnamefont
  {Winkler}},\ and\ \bibinfo {author} {\bibfnamefont {G.}~\bibnamefont
  {Gompper}},\ }\bibfield  {title} {\bibinfo {title} {Physics of
  microswimmers—single particle motion and collective behavior: {A} review},\
  }\href@noop {} {\bibfield  {journal} {\bibinfo  {journal} {Rep. Prog. Phys.}\
  }\textbf {\bibinfo {volume} {78}},\ \bibinfo {pages} {056601} (\bibinfo
  {year} {2015})}\BibitemShut {NoStop}%
\bibitem [{\citenamefont {Hall-Stoodley}\ \emph {et~al.}(2004)\citenamefont
  {Hall-Stoodley}, \citenamefont {Costerton},\ and\ \citenamefont
  {Stoodley}}]{hall2004bacterial}%
  \BibitemOpen
  \bibfield  {author} {\bibinfo {author} {\bibfnamefont {L.}~\bibnamefont
  {Hall-Stoodley}}, \bibinfo {author} {\bibfnamefont {J.~W.}\ \bibnamefont
  {Costerton}},\ and\ \bibinfo {author} {\bibfnamefont {P.}~\bibnamefont
  {Stoodley}},\ }\bibfield  {title} {\bibinfo {title} {Bacterial biofilms:
  {F}rom the natural environment to infectious diseases},\ }\href@noop {}
  {\bibfield  {journal} {\bibinfo  {journal} {Nature Rev. Microbiol.}\ }\textbf
  {\bibinfo {volume} {2}},\ \bibinfo {pages} {95} (\bibinfo {year}
  {2004})}\BibitemShut {NoStop}%
\bibitem [{\citenamefont {Doostmohammadi}\ \emph {et~al.}(2018)\citenamefont
  {Doostmohammadi}, \citenamefont {Ign{\'e}s-Mullol}, \citenamefont {Yeomans},\
  and\ \citenamefont {Sagu{\'e}s}}]{doostmohammadi2018active}%
  \BibitemOpen
  \bibfield  {author} {\bibinfo {author} {\bibfnamefont {A.}~\bibnamefont
  {Doostmohammadi}}, \bibinfo {author} {\bibfnamefont {J.}~\bibnamefont
  {Ign{\'e}s-Mullol}}, \bibinfo {author} {\bibfnamefont {J.~M.}\ \bibnamefont
  {Yeomans}},\ and\ \bibinfo {author} {\bibfnamefont {F.}~\bibnamefont
  {Sagu{\'e}s}},\ }\bibfield  {title} {\bibinfo {title} {Active nematics},\
  }\href@noop {} {\bibfield  {journal} {\bibinfo  {journal} {Nature Commun.}\
  }\textbf {\bibinfo {volume} {9}},\ \bibinfo {pages} {3246} (\bibinfo {year}
  {2018})}\BibitemShut {NoStop}%
\bibitem [{\citenamefont {Be’er}\ and\ \citenamefont
  {Ariel}(2019)}]{be2019statistical}%
  \BibitemOpen
  \bibfield  {author} {\bibinfo {author} {\bibfnamefont {A.}~\bibnamefont
  {Be’er}}\ and\ \bibinfo {author} {\bibfnamefont {G.}~\bibnamefont
  {Ariel}},\ }\bibfield  {title} {\bibinfo {title} {A statistical physics view
  of swarming bacteria},\ }\href@noop {} {\bibfield  {journal} {\bibinfo
  {journal} {Mov. Ecol.}\ }\textbf {\bibinfo {volume} {7}},\ \bibinfo {pages}
  {1} (\bibinfo {year} {2019})}\BibitemShut {NoStop}%
\bibitem [{\citenamefont {Aranson}(2022)}]{aranson2022bacterial}%
  \BibitemOpen
  \bibfield  {author} {\bibinfo {author} {\bibfnamefont {I.~S.}\ \bibnamefont
  {Aranson}},\ }\bibfield  {title} {\bibinfo {title} {Bacterial active
  matter},\ }\href@noop {} {\bibfield  {journal} {\bibinfo  {journal} {Rep.
  Prog. Phys.}\ }\textbf {\bibinfo {volume} {85}},\ \bibinfo {pages} {076601}
  (\bibinfo {year} {2022})}\BibitemShut {NoStop}%
\bibitem [{\citenamefont {Nishiguchi}\ \emph {et~al.}(2018)\citenamefont
  {Nishiguchi}, \citenamefont {Aranson}, \citenamefont {Snezhko},\ and\
  \citenamefont {Sokolov}}]{nishiguchi2018engineering}%
  \BibitemOpen
  \bibfield  {author} {\bibinfo {author} {\bibfnamefont {D.}~\bibnamefont
  {Nishiguchi}}, \bibinfo {author} {\bibfnamefont {I.~S.}\ \bibnamefont
  {Aranson}}, \bibinfo {author} {\bibfnamefont {A.}~\bibnamefont {Snezhko}},\
  and\ \bibinfo {author} {\bibfnamefont {A.}~\bibnamefont {Sokolov}},\
  }\bibfield  {title} {\bibinfo {title} {Engineering bacterial vortex lattice
  via direct laser lithography},\ }\href@noop {} {\bibfield  {journal}
  {\bibinfo  {journal} {Nature Commun.}\ }\textbf {\bibinfo {volume} {9}},\
  \bibinfo {pages} {4486} (\bibinfo {year} {2018})}\BibitemShut {NoStop}%
\bibitem [{\citenamefont {Reinken}\ \emph {et~al.}(2024)\citenamefont
  {Reinken}, \citenamefont {Heidenreich}, \citenamefont {Baer},\ and\
  \citenamefont {Klapp}}]{reinken2024pattern}%
  \BibitemOpen
  \bibfield  {author} {\bibinfo {author} {\bibfnamefont {H.}~\bibnamefont
  {Reinken}}, \bibinfo {author} {\bibfnamefont {S.}~\bibnamefont
  {Heidenreich}}, \bibinfo {author} {\bibfnamefont {M.}~\bibnamefont {Baer}},\
  and\ \bibinfo {author} {\bibfnamefont {S.~H.~L.}\ \bibnamefont {Klapp}},\
  }\bibfield  {title} {\bibinfo {title} {Pattern selection and the route to
  turbulence in incompressible polar active fluids},\ }\href@noop {} {\bibfield
   {journal} {\bibinfo  {journal} {New J. Phys.}\ }\textbf {\bibinfo {volume}
  {26}},\ \bibinfo {pages} {063026} (\bibinfo {year} {2024})}\BibitemShut
  {NoStop}%
\bibitem [{\citenamefont {Dombrowski}\ \emph {et~al.}(2004)\citenamefont
  {Dombrowski}, \citenamefont {Cisneros}, \citenamefont {Chatkaew},
  \citenamefont {Goldstein},\ and\ \citenamefont
  {Kessler}}]{dombrowski2004self}%
  \BibitemOpen
  \bibfield  {author} {\bibinfo {author} {\bibfnamefont {C.}~\bibnamefont
  {Dombrowski}}, \bibinfo {author} {\bibfnamefont {L.}~\bibnamefont
  {Cisneros}}, \bibinfo {author} {\bibfnamefont {S.}~\bibnamefont {Chatkaew}},
  \bibinfo {author} {\bibfnamefont {R.~E.}\ \bibnamefont {Goldstein}},\ and\
  \bibinfo {author} {\bibfnamefont {J.~O.}\ \bibnamefont {Kessler}},\
  }\bibfield  {title} {\bibinfo {title} {Self-concentration and large-scale
  coherence in bacterial dynamics},\ }\href@noop {} {\bibfield  {journal}
  {\bibinfo  {journal} {Phys. Rev. Lett.}\ }\textbf {\bibinfo {volume} {93}},\
  \bibinfo {pages} {098103} (\bibinfo {year} {2004})}\BibitemShut {NoStop}%
\bibitem [{\citenamefont {Wensink}\ \emph {et~al.}(2012)\citenamefont
  {Wensink}, \citenamefont {Dunkel}, \citenamefont {Heidenreich}, \citenamefont
  {Drescher}, \citenamefont {Goldstein}, \citenamefont {L{\"o}wen},\ and\
  \citenamefont {Yeomans}}]{wensink2012meso}%
  \BibitemOpen
  \bibfield  {author} {\bibinfo {author} {\bibfnamefont {H.~H.}\ \bibnamefont
  {Wensink}}, \bibinfo {author} {\bibfnamefont {J.}~\bibnamefont {Dunkel}},
  \bibinfo {author} {\bibfnamefont {S.}~\bibnamefont {Heidenreich}}, \bibinfo
  {author} {\bibfnamefont {K.}~\bibnamefont {Drescher}}, \bibinfo {author}
  {\bibfnamefont {R.~E.}\ \bibnamefont {Goldstein}}, \bibinfo {author}
  {\bibfnamefont {H.}~\bibnamefont {L{\"o}wen}},\ and\ \bibinfo {author}
  {\bibfnamefont {J.~M.}\ \bibnamefont {Yeomans}},\ }\bibfield  {title}
  {\bibinfo {title} {Meso-scale turbulence in living fluids},\ }\href@noop {}
  {\bibfield  {journal} {\bibinfo  {journal} {Proc. Natl. Acad. Sci. U.S.A.}\
  }\textbf {\bibinfo {volume} {109}},\ \bibinfo {pages} {14308} (\bibinfo
  {year} {2012})}\BibitemShut {NoStop}%
\bibitem [{\citenamefont {Reinken}\ \emph {et~al.}(2018)\citenamefont
  {Reinken}, \citenamefont {Klapp}, \citenamefont {B{\"a}r},\ and\
  \citenamefont {Heidenreich}}]{reinken2018derivation}%
  \BibitemOpen
  \bibfield  {author} {\bibinfo {author} {\bibfnamefont {H.}~\bibnamefont
  {Reinken}}, \bibinfo {author} {\bibfnamefont {S.~H.~L.}\ \bibnamefont
  {Klapp}}, \bibinfo {author} {\bibfnamefont {M.}~\bibnamefont {B{\"a}r}},\
  and\ \bibinfo {author} {\bibfnamefont {S.}~\bibnamefont {Heidenreich}},\
  }\bibfield  {title} {\bibinfo {title} {Derivation of a hydrodynamic theory
  for mesoscale dynamics in microswimmer suspensions},\ }\href@noop {}
  {\bibfield  {journal} {\bibinfo  {journal} {Phys. Rev. E}\ }\textbf {\bibinfo
  {volume} {97}},\ \bibinfo {pages} {022613} (\bibinfo {year}
  {2018})}\BibitemShut {NoStop}%
\bibitem [{\citenamefont {Alert}\ \emph {et~al.}(2022)\citenamefont {Alert},
  \citenamefont {Casademunt},\ and\ \citenamefont {Joanny}}]{alert2021active}%
  \BibitemOpen
  \bibfield  {author} {\bibinfo {author} {\bibfnamefont {R.}~\bibnamefont
  {Alert}}, \bibinfo {author} {\bibfnamefont {J.}~\bibnamefont {Casademunt}},\
  and\ \bibinfo {author} {\bibfnamefont {J.-F.}\ \bibnamefont {Joanny}},\
  }\bibfield  {title} {\bibinfo {title} {Active turbulence},\ }\href@noop {}
  {\bibfield  {journal} {\bibinfo  {journal} {Annu. Rev. Condens. Matter
  Phys.}\ }\textbf {\bibinfo {volume} {13}} (\bibinfo {year}
  {2022})}\BibitemShut {NoStop}%
\bibitem [{\citenamefont {Jeckel}\ \emph {et~al.}(2019)\citenamefont {Jeckel},
  \citenamefont {Jelli}, \citenamefont {Hartmann}, \citenamefont {Singh},
  \citenamefont {Mok}, \citenamefont {Totz}, \citenamefont {Vidakovic},
  \citenamefont {Eckhardt}, \citenamefont {Dunkel},\ and\ \citenamefont
  {Drescher}}]{jeckel2019learning}%
  \BibitemOpen
  \bibfield  {author} {\bibinfo {author} {\bibfnamefont {H.}~\bibnamefont
  {Jeckel}}, \bibinfo {author} {\bibfnamefont {E.}~\bibnamefont {Jelli}},
  \bibinfo {author} {\bibfnamefont {R.}~\bibnamefont {Hartmann}}, \bibinfo
  {author} {\bibfnamefont {P.~K.}\ \bibnamefont {Singh}}, \bibinfo {author}
  {\bibfnamefont {R.}~\bibnamefont {Mok}}, \bibinfo {author} {\bibfnamefont
  {J.~F.}\ \bibnamefont {Totz}}, \bibinfo {author} {\bibfnamefont
  {L.}~\bibnamefont {Vidakovic}}, \bibinfo {author} {\bibfnamefont
  {B.}~\bibnamefont {Eckhardt}}, \bibinfo {author} {\bibfnamefont
  {J.}~\bibnamefont {Dunkel}},\ and\ \bibinfo {author} {\bibfnamefont
  {K.}~\bibnamefont {Drescher}},\ }\bibfield  {title} {\bibinfo {title}
  {Learning the space-time phase diagram of bacterial swarm expansion},\
  }\href@noop {} {\bibfield  {journal} {\bibinfo  {journal} {Proc. Natl. Acad.
  Sci. U.S.A.}\ }\textbf {\bibinfo {volume} {116}},\ \bibinfo {pages} {1489}
  (\bibinfo {year} {2019})}\BibitemShut {NoStop}%
\bibitem [{\citenamefont {Be’er}\ \emph {et~al.}(2020)\citenamefont
  {Be’er}, \citenamefont {Ilkanaiv}, \citenamefont {Gross}, \citenamefont
  {Kearns}, \citenamefont {Heidenreich}, \citenamefont {B{\"a}r},\ and\
  \citenamefont {Ariel}}]{be2020phase}%
  \BibitemOpen
  \bibfield  {author} {\bibinfo {author} {\bibfnamefont {A.}~\bibnamefont
  {Be’er}}, \bibinfo {author} {\bibfnamefont {B.}~\bibnamefont {Ilkanaiv}},
  \bibinfo {author} {\bibfnamefont {R.}~\bibnamefont {Gross}}, \bibinfo
  {author} {\bibfnamefont {D.~B.}\ \bibnamefont {Kearns}}, \bibinfo {author}
  {\bibfnamefont {S.}~\bibnamefont {Heidenreich}}, \bibinfo {author}
  {\bibfnamefont {M.}~\bibnamefont {B{\"a}r}},\ and\ \bibinfo {author}
  {\bibfnamefont {G.}~\bibnamefont {Ariel}},\ }\bibfield  {title} {\bibinfo
  {title} {A phase diagram for bacterial swarming},\ }\href@noop {} {\bibfield
  {journal} {\bibinfo  {journal} {Commun. Phys.}\ }\textbf {\bibinfo {volume}
  {3}},\ \bibinfo {pages} {66} (\bibinfo {year} {2020})}\BibitemShut {NoStop}%
\bibitem [{\citenamefont {Vicsek}\ \emph {et~al.}(1995)\citenamefont {Vicsek},
  \citenamefont {Czir{\'o}k}, \citenamefont {Ben-Jacob}, \citenamefont
  {Cohen},\ and\ \citenamefont {Shochet}}]{vicsek1995novel}%
  \BibitemOpen
  \bibfield  {author} {\bibinfo {author} {\bibfnamefont {T.}~\bibnamefont
  {Vicsek}}, \bibinfo {author} {\bibfnamefont {A.}~\bibnamefont {Czir{\'o}k}},
  \bibinfo {author} {\bibfnamefont {E.}~\bibnamefont {Ben-Jacob}}, \bibinfo
  {author} {\bibfnamefont {I.}~\bibnamefont {Cohen}},\ and\ \bibinfo {author}
  {\bibfnamefont {O.}~\bibnamefont {Shochet}},\ }\bibfield  {title} {\bibinfo
  {title} {Novel type of phase transition in a system of self-driven
  particles},\ }\href@noop {} {\bibfield  {journal} {\bibinfo  {journal} {Phys.
  Rev. Lett.}\ }\textbf {\bibinfo {volume} {75}},\ \bibinfo {pages} {1226}
  (\bibinfo {year} {1995})}\BibitemShut {NoStop}%
\bibitem [{\citenamefont {Toner}\ and\ \citenamefont
  {Tu}(1998)}]{toner1998flocks}%
  \BibitemOpen
  \bibfield  {author} {\bibinfo {author} {\bibfnamefont {J.}~\bibnamefont
  {Toner}}\ and\ \bibinfo {author} {\bibfnamefont {Y.}~\bibnamefont {Tu}},\
  }\bibfield  {title} {\bibinfo {title} {Flocks, herds, and schools: A
  quantitative theory of flocking},\ }\href@noop {} {\bibfield  {journal}
  {\bibinfo  {journal} {Phys. Rev. E}\ }\textbf {\bibinfo {volume} {58}},\
  \bibinfo {pages} {4828} (\bibinfo {year} {1998})}\BibitemShut {NoStop}%
\bibitem [{\citenamefont {Toner}\ \emph {et~al.}(2005)\citenamefont {Toner},
  \citenamefont {Tu},\ and\ \citenamefont
  {Ramaswamy}}]{toner2005hydrodynamics}%
  \BibitemOpen
  \bibfield  {author} {\bibinfo {author} {\bibfnamefont {J.}~\bibnamefont
  {Toner}}, \bibinfo {author} {\bibfnamefont {Y.}~\bibnamefont {Tu}},\ and\
  \bibinfo {author} {\bibfnamefont {S.}~\bibnamefont {Ramaswamy}},\ }\bibfield
  {title} {\bibinfo {title} {Hydrodynamics and phases of flocks},\ }\href@noop
  {} {\bibfield  {journal} {\bibinfo  {journal} {Ann. Phys.}\ }\textbf
  {\bibinfo {volume} {318}},\ \bibinfo {pages} {170} (\bibinfo {year}
  {2005})}\BibitemShut {NoStop}%
\bibitem [{\citenamefont {Shen}\ \emph {et~al.}(2016)\citenamefont {Shen},
  \citenamefont {Tan},\ and\ \citenamefont {Xu}}]{shen2016probing}%
  \BibitemOpen
  \bibfield  {author} {\bibinfo {author} {\bibfnamefont {H.}~\bibnamefont
  {Shen}}, \bibinfo {author} {\bibfnamefont {P.}~\bibnamefont {Tan}},\ and\
  \bibinfo {author} {\bibfnamefont {L.}~\bibnamefont {Xu}},\ }\bibfield
  {title} {\bibinfo {title} {Probing the role of mobility in the collective
  motion of nonequilibrium systems},\ }\href@noop {} {\bibfield  {journal}
  {\bibinfo  {journal} {Phys. Rev. Lett.}\ }\textbf {\bibinfo {volume} {116}},\
  \bibinfo {pages} {048302} (\bibinfo {year} {2016})}\BibitemShut {NoStop}%
\bibitem [{\citenamefont {Ferrante}\ \emph {et~al.}(2013)\citenamefont
  {Ferrante}, \citenamefont {Turgut}, \citenamefont {Dorigo},\ and\
  \citenamefont {Huepe}}]{ferrante2013elasticity}%
  \BibitemOpen
  \bibfield  {author} {\bibinfo {author} {\bibfnamefont {E.}~\bibnamefont
  {Ferrante}}, \bibinfo {author} {\bibfnamefont {A.~E.}\ \bibnamefont
  {Turgut}}, \bibinfo {author} {\bibfnamefont {M.}~\bibnamefont {Dorigo}},\
  and\ \bibinfo {author} {\bibfnamefont {C.}~\bibnamefont {Huepe}},\ }\bibfield
   {title} {\bibinfo {title} {Elasticity-based mechanism for the collective
  motion of self-propelled particles with springlike interactions: {A} model
  system for natural and artificial swarms},\ }\href@noop {} {\bibfield
  {journal} {\bibinfo  {journal} {Phys. Rev. Lett.}\ }\textbf {\bibinfo
  {volume} {111}},\ \bibinfo {pages} {268302} (\bibinfo {year}
  {2013})}\BibitemShut {NoStop}%
\bibitem [{\citenamefont {Hern{\'a}ndez-L{\'o}pez}\ \emph
  {et~al.}(2024)\citenamefont {Hern{\'a}ndez-L{\'o}pez}, \citenamefont
  {Baconnier}, \citenamefont {Coulais}, \citenamefont {Dauchot},\ and\
  \citenamefont {D{\"u}ring}}]{hernandez2024model}%
  \BibitemOpen
  \bibfield  {author} {\bibinfo {author} {\bibfnamefont {C.}~\bibnamefont
  {Hern{\'a}ndez-L{\'o}pez}}, \bibinfo {author} {\bibfnamefont
  {P.}~\bibnamefont {Baconnier}}, \bibinfo {author} {\bibfnamefont
  {C.}~\bibnamefont {Coulais}}, \bibinfo {author} {\bibfnamefont
  {O.}~\bibnamefont {Dauchot}},\ and\ \bibinfo {author} {\bibfnamefont
  {G.}~\bibnamefont {D{\"u}ring}},\ }\bibfield  {title} {\bibinfo {title}
  {Model of active solids: {R}igid body motion and shape-changing mechanisms},\
  }\href@noop {} {\bibfield  {journal} {\bibinfo  {journal} {Phys. Rev. Lett.}\
  }\textbf {\bibinfo {volume} {132}},\ \bibinfo {pages} {238303} (\bibinfo
  {year} {2024})}\BibitemShut {NoStop}%
\bibitem [{\citenamefont {Baconnier}\ \emph {et~al.}(2024)\citenamefont
  {Baconnier}, \citenamefont {D{\'e}mery},\ and\ \citenamefont
  {Dauchot}}]{baconnier2024noise}%
  \BibitemOpen
  \bibfield  {author} {\bibinfo {author} {\bibfnamefont {P.}~\bibnamefont
  {Baconnier}}, \bibinfo {author} {\bibfnamefont {V.}~\bibnamefont
  {D{\'e}mery}},\ and\ \bibinfo {author} {\bibfnamefont {O.}~\bibnamefont
  {Dauchot}},\ }\bibfield  {title} {\bibinfo {title} {Noise-induced collective
  actuation in active solids},\ }\href@noop {} {\bibfield  {journal} {\bibinfo
  {journal} {Phys. Rev. E}\ }\textbf {\bibinfo {volume} {109}},\ \bibinfo
  {pages} {024606} (\bibinfo {year} {2024})}\BibitemShut {NoStop}%
\bibitem [{\citenamefont {Hawkins}\ and\ \citenamefont
  {Liverpool}(2014)}]{hawkins2014stress}%
  \BibitemOpen
  \bibfield  {author} {\bibinfo {author} {\bibfnamefont {R.~J.}\ \bibnamefont
  {Hawkins}}\ and\ \bibinfo {author} {\bibfnamefont {T.~B.}\ \bibnamefont
  {Liverpool}},\ }\bibfield  {title} {\bibinfo {title} {Stress reorganization
  and response in active solids},\ }\href@noop {} {\bibfield  {journal}
  {\bibinfo  {journal} {Phys. Rev. Lett.}\ }\textbf {\bibinfo {volume} {113}},\
  \bibinfo {pages} {028102} (\bibinfo {year} {2014})}\BibitemShut {NoStop}%
\bibitem [{\citenamefont {Maitra}\ and\ \citenamefont
  {Ramaswamy}(2019)}]{maitra2019oriented}%
  \BibitemOpen
  \bibfield  {author} {\bibinfo {author} {\bibfnamefont {A.}~\bibnamefont
  {Maitra}}\ and\ \bibinfo {author} {\bibfnamefont {S.}~\bibnamefont
  {Ramaswamy}},\ }\bibfield  {title} {\bibinfo {title} {Oriented active
  solids},\ }\href@noop {} {\bibfield  {journal} {\bibinfo  {journal} {Phys.
  Rev. Lett.}\ }\textbf {\bibinfo {volume} {123}},\ \bibinfo {pages} {238001}
  (\bibinfo {year} {2019})}\BibitemShut {NoStop}%
\bibitem [{\citenamefont {Needleman}\ and\ \citenamefont
  {Dogic}(2017)}]{needleman2017active}%
  \BibitemOpen
  \bibfield  {author} {\bibinfo {author} {\bibfnamefont {D.}~\bibnamefont
  {Needleman}}\ and\ \bibinfo {author} {\bibfnamefont {Z.}~\bibnamefont
  {Dogic}},\ }\bibfield  {title} {\bibinfo {title} {Active matter at the
  interface between materials science and cell biology},\ }\href@noop {}
  {\bibfield  {journal} {\bibinfo  {journal} {Nature Rev. Mater.}\ }\textbf
  {\bibinfo {volume} {2}} (\bibinfo {year} {2017})}\BibitemShut {NoStop}%
\bibitem [{\citenamefont {Baconnier}\ \emph {et~al.}(2022)\citenamefont
  {Baconnier}, \citenamefont {Shohat}, \citenamefont {L{\'o}pez}, \citenamefont
  {Coulais}, \citenamefont {D{\'e}mery}, \citenamefont {D{\"u}ring},\ and\
  \citenamefont {Dauchot}}]{baconnier2022selective}%
  \BibitemOpen
  \bibfield  {author} {\bibinfo {author} {\bibfnamefont {P.}~\bibnamefont
  {Baconnier}}, \bibinfo {author} {\bibfnamefont {D.}~\bibnamefont {Shohat}},
  \bibinfo {author} {\bibfnamefont {C.~H.}\ \bibnamefont {L{\'o}pez}}, \bibinfo
  {author} {\bibfnamefont {C.}~\bibnamefont {Coulais}}, \bibinfo {author}
  {\bibfnamefont {V.}~\bibnamefont {D{\'e}mery}}, \bibinfo {author}
  {\bibfnamefont {G.}~\bibnamefont {D{\"u}ring}},\ and\ \bibinfo {author}
  {\bibfnamefont {O.}~\bibnamefont {Dauchot}},\ }\bibfield  {title} {\bibinfo
  {title} {Selective and collective actuation in active solids},\ }\href@noop
  {} {\bibfield  {journal} {\bibinfo  {journal} {Nature Phys.}\ }\textbf
  {\bibinfo {volume} {18}},\ \bibinfo {pages} {1234} (\bibinfo {year}
  {2022})}\BibitemShut {NoStop}%
\bibitem [{\citenamefont {Caprini}\ \emph {et~al.}(2023)\citenamefont
  {Caprini}, \citenamefont {Marini Bettolo~Marconi}, \citenamefont {Puglisi},\
  and\ \citenamefont {L{\"o}wen}}]{caprini2023entropons}%
  \BibitemOpen
  \bibfield  {author} {\bibinfo {author} {\bibfnamefont {L.}~\bibnamefont
  {Caprini}}, \bibinfo {author} {\bibfnamefont {U.}~\bibnamefont {Marini
  Bettolo~Marconi}}, \bibinfo {author} {\bibfnamefont {A.}~\bibnamefont
  {Puglisi}},\ and\ \bibinfo {author} {\bibfnamefont {H.}~\bibnamefont
  {L{\"o}wen}},\ }\bibfield  {title} {\bibinfo {title} {Entropons as collective
  excitations in active solids},\ }\href@noop {} {\bibfield  {journal}
  {\bibinfo  {journal} {J. Chem. Phys.}\ }\textbf {\bibinfo {volume} {159}}
  (\bibinfo {year} {2023})}\BibitemShut {NoStop}%
\bibitem [{\citenamefont {Kinoshita}\ \emph {et~al.}(2025)\citenamefont
  {Kinoshita}, \citenamefont {Uchida},\ and\ \citenamefont
  {Menzel}}]{kinoshita2025collective}%
  \BibitemOpen
  \bibfield  {author} {\bibinfo {author} {\bibfnamefont {Y.}~\bibnamefont
  {Kinoshita}}, \bibinfo {author} {\bibfnamefont {N.}~\bibnamefont {Uchida}},\
  and\ \bibinfo {author} {\bibfnamefont {A.~M.}\ \bibnamefont {Menzel}},\
  }\bibfield  {title} {\bibinfo {title} {Collective excitations in active
  solids featuring alignment interactions},\ }\href@noop {} {\bibfield
  {journal} {\bibinfo  {journal} {J. Chemi. Phys.}\ }\textbf {\bibinfo {volume}
  {162}} (\bibinfo {year} {2025})}\BibitemShut {NoStop}%
\bibitem [{\citenamefont {L{\aa}ng}\ \emph {et~al.}(2024)\citenamefont
  {L{\aa}ng}, \citenamefont {L{\aa}ng}, \citenamefont {Blicher}, \citenamefont
  {Rognes}, \citenamefont {Dommersnes},\ and\ \citenamefont
  {B{\o}e}}]{laang2024topology}%
  \BibitemOpen
  \bibfield  {author} {\bibinfo {author} {\bibfnamefont {E.}~\bibnamefont
  {L{\aa}ng}}, \bibinfo {author} {\bibfnamefont {A.}~\bibnamefont {L{\aa}ng}},
  \bibinfo {author} {\bibfnamefont {P.}~\bibnamefont {Blicher}}, \bibinfo
  {author} {\bibfnamefont {T.}~\bibnamefont {Rognes}}, \bibinfo {author}
  {\bibfnamefont {P.~G.}\ \bibnamefont {Dommersnes}},\ and\ \bibinfo {author}
  {\bibfnamefont {S.~O.}\ \bibnamefont {B{\o}e}},\ }\bibfield  {title}
  {\bibinfo {title} {Topology-guided polar ordering of collective cell
  migration},\ }\href@noop {} {\bibfield  {journal} {\bibinfo  {journal} {Sci.
  Adv.}\ }\textbf {\bibinfo {volume} {10}},\ \bibinfo {pages} {eadk4825}
  (\bibinfo {year} {2024})}\BibitemShut {NoStop}%
\bibitem [{\citenamefont {Scheibner}\ \emph {et~al.}(2020)\citenamefont
  {Scheibner}, \citenamefont {Souslov}, \citenamefont {Banerjee}, \citenamefont
  {Sur{\'o}wka}, \citenamefont {Irvine},\ and\ \citenamefont
  {Vitelli}}]{scheibner2020odd}%
  \BibitemOpen
  \bibfield  {author} {\bibinfo {author} {\bibfnamefont {C.}~\bibnamefont
  {Scheibner}}, \bibinfo {author} {\bibfnamefont {A.}~\bibnamefont {Souslov}},
  \bibinfo {author} {\bibfnamefont {D.}~\bibnamefont {Banerjee}}, \bibinfo
  {author} {\bibfnamefont {P.}~\bibnamefont {Sur{\'o}wka}}, \bibinfo {author}
  {\bibfnamefont {W.~T.}\ \bibnamefont {Irvine}},\ and\ \bibinfo {author}
  {\bibfnamefont {V.}~\bibnamefont {Vitelli}},\ }\bibfield  {title} {\bibinfo
  {title} {Odd elasticity},\ }\href@noop {} {\bibfield  {journal} {\bibinfo
  {journal} {Nature Phys.}\ }\textbf {\bibinfo {volume} {16}},\ \bibinfo
  {pages} {475} (\bibinfo {year} {2020})}\BibitemShut {NoStop}%
\bibitem [{\citenamefont {Fruchart}\ \emph {et~al.}(2023)\citenamefont
  {Fruchart}, \citenamefont {Scheibner},\ and\ \citenamefont
  {Vitelli}}]{fruchart2023odd}%
  \BibitemOpen
  \bibfield  {author} {\bibinfo {author} {\bibfnamefont {M.}~\bibnamefont
  {Fruchart}}, \bibinfo {author} {\bibfnamefont {C.}~\bibnamefont
  {Scheibner}},\ and\ \bibinfo {author} {\bibfnamefont {V.}~\bibnamefont
  {Vitelli}},\ }\bibfield  {title} {\bibinfo {title} {Odd viscosity and odd
  elasticity},\ }\href@noop {} {\bibfield  {journal} {\bibinfo  {journal}
  {Annu. Rev. Condens. Matter Phys.}\ }\textbf {\bibinfo {volume} {14}},\
  \bibinfo {pages} {471} (\bibinfo {year} {2023})}\BibitemShut {NoStop}%
\bibitem [{\citenamefont {Saintillan}\ and\ \citenamefont
  {Shelley}(2013)}]{saintillan2013active}%
  \BibitemOpen
  \bibfield  {author} {\bibinfo {author} {\bibfnamefont {D.}~\bibnamefont
  {Saintillan}}\ and\ \bibinfo {author} {\bibfnamefont {M.~J.}\ \bibnamefont
  {Shelley}},\ }\bibfield  {title} {\bibinfo {title} {Active suspensions and
  their nonlinear models},\ }\href@noop {} {\bibfield  {journal} {\bibinfo
  {journal} {C. R. Phys.}\ }\textbf {\bibinfo {volume} {14}},\ \bibinfo {pages}
  {497} (\bibinfo {year} {2013})}\BibitemShut {NoStop}%
\bibitem [{\citenamefont {Hemingway}\ \emph {et~al.}(2015)\citenamefont
  {Hemingway}, \citenamefont {Maitra}, \citenamefont {Banerjee}, \citenamefont
  {Marchetti}, \citenamefont {Ramaswamy}, \citenamefont {Fielding},\ and\
  \citenamefont {Cates}}]{hemingway2015active}%
  \BibitemOpen
  \bibfield  {author} {\bibinfo {author} {\bibfnamefont {E.~J.}\ \bibnamefont
  {Hemingway}}, \bibinfo {author} {\bibfnamefont {A.}~\bibnamefont {Maitra}},
  \bibinfo {author} {\bibfnamefont {S.}~\bibnamefont {Banerjee}}, \bibinfo
  {author} {\bibfnamefont {M.~C.}\ \bibnamefont {Marchetti}}, \bibinfo {author}
  {\bibfnamefont {S.}~\bibnamefont {Ramaswamy}}, \bibinfo {author}
  {\bibfnamefont {S.~M.}\ \bibnamefont {Fielding}},\ and\ \bibinfo {author}
  {\bibfnamefont {M.~E.}\ \bibnamefont {Cates}},\ }\bibfield  {title} {\bibinfo
  {title} {Active viscoelastic matter: {F}rom bacterial drag reduction to
  turbulent solids},\ }\href@noop {} {\bibfield  {journal} {\bibinfo  {journal}
  {Phys. Rev. Lett.}\ }\textbf {\bibinfo {volume} {114}},\ \bibinfo {pages}
  {098302} (\bibinfo {year} {2015})}\BibitemShut {NoStop}%
\bibitem [{\citenamefont {Hemingway}\ \emph {et~al.}(2016)\citenamefont
  {Hemingway}, \citenamefont {Cates},\ and\ \citenamefont
  {Fielding}}]{hemingway2016viscoelastic}%
  \BibitemOpen
  \bibfield  {author} {\bibinfo {author} {\bibfnamefont {E.~J.}\ \bibnamefont
  {Hemingway}}, \bibinfo {author} {\bibfnamefont {M.~E.}\ \bibnamefont
  {Cates}},\ and\ \bibinfo {author} {\bibfnamefont {S.~M.}\ \bibnamefont
  {Fielding}},\ }\bibfield  {title} {\bibinfo {title} {Viscoelastic and
  elastomeric active matter: {L}inear instability and nonlinear dynamics},\
  }\href@noop {} {\bibfield  {journal} {\bibinfo  {journal} {Phys. Rev. E}\
  }\textbf {\bibinfo {volume} {93}},\ \bibinfo {pages} {032702} (\bibinfo
  {year} {2016})}\BibitemShut {NoStop}%
\bibitem [{\citenamefont {{E. L. C. VI M. Plan}}\ \emph
  {et~al.}(2020)\citenamefont {{E. L. C. VI M. Plan}}, \citenamefont
  {Yeomans},\ and\ \citenamefont {Doostmohammadi}}]{plan2020active}%
  \BibitemOpen
  \bibfield  {author} {\bibinfo {author} {\bibnamefont {{E. L. C. VI M.
  Plan}}}, \bibinfo {author} {\bibfnamefont {J.~M.}\ \bibnamefont {Yeomans}},\
  and\ \bibinfo {author} {\bibfnamefont {A.}~\bibnamefont {Doostmohammadi}},\
  }\bibfield  {title} {\bibinfo {title} {Active matter in a viscoelastic
  environment},\ }\href@noop {} {\bibfield  {journal} {\bibinfo  {journal}
  {Phys. Rev. Fluids}\ }\textbf {\bibinfo {volume} {5}},\ \bibinfo {pages}
  {023102} (\bibinfo {year} {2020})}\BibitemShut {NoStop}%
\bibitem [{\citenamefont {Reinken}\ and\ \citenamefont
  {Menzel}(2024)}]{reinken2024vortex}%
  \BibitemOpen
  \bibfield  {author} {\bibinfo {author} {\bibfnamefont {H.}~\bibnamefont
  {Reinken}}\ and\ \bibinfo {author} {\bibfnamefont {A.~M.}\ \bibnamefont
  {Menzel}},\ }\bibfield  {title} {\bibinfo {title} {Vortex pattern
  stabilization in thin films resulting from shear thickening of active
  suspensions},\ }\href@noop {} {\bibfield  {journal} {\bibinfo  {journal}
  {Phys. Rev. Lett.}\ }\textbf {\bibinfo {volume} {132}},\ \bibinfo {pages}
  {138301} (\bibinfo {year} {2024})}\BibitemShut {NoStop}%
\bibitem [{\citenamefont {Reinken}\ and\ \citenamefont
  {Menzel}(2025{\natexlab{a}})}]{reinken2025self}%
  \BibitemOpen
  \bibfield  {author} {\bibinfo {author} {\bibfnamefont {H.}~\bibnamefont
  {Reinken}}\ and\ \bibinfo {author} {\bibfnamefont {A.~M.}\ \bibnamefont
  {Menzel}},\ }\bibfield  {title} {\bibinfo {title} {Self-sustained patchy
  turbulence in shear-thinning active fluids},\ }\href@noop {} {\bibfield
  {journal} {\bibinfo  {journal} {Commun. Phys.}\ }\textbf {\bibinfo {volume}
  {8}},\ \bibinfo {pages} {270} (\bibinfo {year}
  {2025}{\natexlab{a}})}\BibitemShut {NoStop}%
\bibitem [{\citenamefont {K{\"o}pf}\ and\ \citenamefont
  {Pismen}(2013)}]{kopf2013non}%
  \BibitemOpen
  \bibfield  {author} {\bibinfo {author} {\bibfnamefont {M.~H.}\ \bibnamefont
  {K{\"o}pf}}\ and\ \bibinfo {author} {\bibfnamefont {L.~M.}\ \bibnamefont
  {Pismen}},\ }\bibfield  {title} {\bibinfo {title} {Non-equilibrium patterns
  in polarizable active layers},\ }\href@noop {} {\bibfield  {journal}
  {\bibinfo  {journal} {Physica D}\ }\textbf {\bibinfo {volume} {259}},\
  \bibinfo {pages} {48} (\bibinfo {year} {2013})}\BibitemShut {NoStop}%
\bibitem [{\citenamefont {Choudhary}\ \emph {et~al.}(2023)\citenamefont
  {Choudhary}, \citenamefont {Nambiar},\ and\ \citenamefont
  {Stark}}]{choudhary2023orientational}%
  \BibitemOpen
  \bibfield  {author} {\bibinfo {author} {\bibfnamefont {A.}~\bibnamefont
  {Choudhary}}, \bibinfo {author} {\bibfnamefont {S.}~\bibnamefont {Nambiar}},\
  and\ \bibinfo {author} {\bibfnamefont {H.}~\bibnamefont {Stark}},\ }\bibfield
   {title} {\bibinfo {title} {Orientational dynamics and rheology of active
  suspensions in weakly viscoelastic flows},\ }\href@noop {} {\bibfield
  {journal} {\bibinfo  {journal} {Commun. Phys.}\ }\textbf {\bibinfo {volume}
  {6}},\ \bibinfo {pages} {163} (\bibinfo {year} {2023})}\BibitemShut {NoStop}%
\bibitem [{\citenamefont {Mathijssen}\ \emph {et~al.}(2016)\citenamefont
  {Mathijssen}, \citenamefont {Shendruk}, \citenamefont {Yeomans},\ and\
  \citenamefont {Doostmohammadi}}]{mathijssen2016upstream}%
  \BibitemOpen
  \bibfield  {author} {\bibinfo {author} {\bibfnamefont {A.~J.}\ \bibnamefont
  {Mathijssen}}, \bibinfo {author} {\bibfnamefont {T.~N.}\ \bibnamefont
  {Shendruk}}, \bibinfo {author} {\bibfnamefont {J.~M.}\ \bibnamefont
  {Yeomans}},\ and\ \bibinfo {author} {\bibfnamefont {A.}~\bibnamefont
  {Doostmohammadi}},\ }\bibfield  {title} {\bibinfo {title} {Upstream swimming
  in microbiological flows},\ }\href@noop {} {\bibfield  {journal} {\bibinfo
  {journal} {Phys. Rev. Lett.}\ }\textbf {\bibinfo {volume} {116}},\ \bibinfo
  {pages} {028104} (\bibinfo {year} {2016})}\BibitemShut {NoStop}%
\bibitem [{\citenamefont {Joanny}\ \emph {et~al.}(2007)\citenamefont {Joanny},
  \citenamefont {J{\"u}licher}, \citenamefont {Kruse},\ and\ \citenamefont
  {Prost}}]{joanny2007hydrodynamic}%
  \BibitemOpen
  \bibfield  {author} {\bibinfo {author} {\bibfnamefont {J.-F.}\ \bibnamefont
  {Joanny}}, \bibinfo {author} {\bibfnamefont {F.}~\bibnamefont
  {J{\"u}licher}}, \bibinfo {author} {\bibfnamefont {K.}~\bibnamefont
  {Kruse}},\ and\ \bibinfo {author} {\bibfnamefont {J.}~\bibnamefont {Prost}},\
  }\bibfield  {title} {\bibinfo {title} {Hydrodynamic theory for
  multi-component active polar gels},\ }\href@noop {} {\bibfield  {journal}
  {\bibinfo  {journal} {New J. Phys.}\ }\textbf {\bibinfo {volume} {9}},\
  \bibinfo {pages} {422} (\bibinfo {year} {2007})}\BibitemShut {NoStop}%
\bibitem [{\citenamefont {Duclut}\ \emph {et~al.}(2024)\citenamefont {Duclut},
  \citenamefont {Bo}, \citenamefont {Lier}, \citenamefont {Armas},
  \citenamefont {Sur{\'o}wka},\ and\ \citenamefont
  {J{\"u}licher}}]{duclut2024probe}%
  \BibitemOpen
  \bibfield  {author} {\bibinfo {author} {\bibfnamefont {C.}~\bibnamefont
  {Duclut}}, \bibinfo {author} {\bibfnamefont {S.}~\bibnamefont {Bo}}, \bibinfo
  {author} {\bibfnamefont {R.}~\bibnamefont {Lier}}, \bibinfo {author}
  {\bibfnamefont {J.}~\bibnamefont {Armas}}, \bibinfo {author} {\bibfnamefont
  {P.}~\bibnamefont {Sur{\'o}wka}},\ and\ \bibinfo {author} {\bibfnamefont
  {F.}~\bibnamefont {J{\"u}licher}},\ }\bibfield  {title} {\bibinfo {title}
  {Probe particles in odd active viscoelastic fluids: How activity and
  dissipation determine linear stability},\ }\href@noop {} {\bibfield
  {journal} {\bibinfo  {journal} {Phys. Rev. E}\ }\textbf {\bibinfo {volume}
  {109}},\ \bibinfo {pages} {044126} (\bibinfo {year} {2024})}\BibitemShut
  {NoStop}%
\bibitem [{\citenamefont {Li}\ \emph {et~al.}(2021)\citenamefont {Li},
  \citenamefont {Lauga},\ and\ \citenamefont {Ardekani}}]{li2021microswimming}%
  \BibitemOpen
  \bibfield  {author} {\bibinfo {author} {\bibfnamefont {G.}~\bibnamefont
  {Li}}, \bibinfo {author} {\bibfnamefont {E.}~\bibnamefont {Lauga}},\ and\
  \bibinfo {author} {\bibfnamefont {A.~M.}\ \bibnamefont {Ardekani}},\
  }\bibfield  {title} {\bibinfo {title} {Microswimming in viscoelastic
  fluids},\ }\href@noop {} {\bibfield  {journal} {\bibinfo  {journal} {J.
  Non-{N}ewt. Fluid Mech.}\ }\textbf {\bibinfo {volume} {297}},\ \bibinfo
  {pages} {104655} (\bibinfo {year} {2021})}\BibitemShut {NoStop}%
\bibitem [{\citenamefont {Jana}\ \emph {et~al.}(2020)\citenamefont {Jana},
  \citenamefont {Charlton}, \citenamefont {Eland}, \citenamefont {Burgess},
  \citenamefont {Wipat}, \citenamefont {Curtis},\ and\ \citenamefont
  {Chen}}]{jana2020nonlinear}%
  \BibitemOpen
  \bibfield  {author} {\bibinfo {author} {\bibfnamefont {S.}~\bibnamefont
  {Jana}}, \bibinfo {author} {\bibfnamefont {S.~G.}\ \bibnamefont {Charlton}},
  \bibinfo {author} {\bibfnamefont {L.~E.}\ \bibnamefont {Eland}}, \bibinfo
  {author} {\bibfnamefont {J.~G.}\ \bibnamefont {Burgess}}, \bibinfo {author}
  {\bibfnamefont {A.}~\bibnamefont {Wipat}}, \bibinfo {author} {\bibfnamefont
  {T.~P.}\ \bibnamefont {Curtis}},\ and\ \bibinfo {author} {\bibfnamefont
  {J.}~\bibnamefont {Chen}},\ }\bibfield  {title} {\bibinfo {title} {Nonlinear
  rheological characteristics of single species bacterial biofilms},\
  }\href@noop {} {\bibfield  {journal} {\bibinfo  {journal} {npj Biofilms
  Microbiomes}\ }\textbf {\bibinfo {volume} {6}},\ \bibinfo {pages} {19}
  (\bibinfo {year} {2020})}\BibitemShut {NoStop}%
\bibitem [{\citenamefont {Worlitzer}\ \emph {et~al.}(2022)\citenamefont
  {Worlitzer}, \citenamefont {Jose}, \citenamefont {Grinberg}, \citenamefont
  {B{\"a}r}, \citenamefont {Heidenreich}, \citenamefont {Eldar}, \citenamefont
  {Ariel},\ and\ \citenamefont {Be’er}}]{worlitzer2022biophysical}%
  \BibitemOpen
  \bibfield  {author} {\bibinfo {author} {\bibfnamefont {V.~M.}\ \bibnamefont
  {Worlitzer}}, \bibinfo {author} {\bibfnamefont {A.}~\bibnamefont {Jose}},
  \bibinfo {author} {\bibfnamefont {I.}~\bibnamefont {Grinberg}}, \bibinfo
  {author} {\bibfnamefont {M.}~\bibnamefont {B{\"a}r}}, \bibinfo {author}
  {\bibfnamefont {S.}~\bibnamefont {Heidenreich}}, \bibinfo {author}
  {\bibfnamefont {A.}~\bibnamefont {Eldar}}, \bibinfo {author} {\bibfnamefont
  {G.}~\bibnamefont {Ariel}},\ and\ \bibinfo {author} {\bibfnamefont
  {A.}~\bibnamefont {Be’er}},\ }\bibfield  {title} {\bibinfo {title}
  {Biophysical aspects underlying the swarm to biofilm transition},\
  }\href@noop {} {\bibfield  {journal} {\bibinfo  {journal} {Sci. Adv.}\
  }\textbf {\bibinfo {volume} {8}},\ \bibinfo {pages} {eabn8152} (\bibinfo
  {year} {2022})}\BibitemShut {NoStop}%
\bibitem [{\citenamefont {Reinken}\ and\ \citenamefont
  {Menzel}(2025{\natexlab{b}})}]{reinken2025rheologically}%
  \BibitemOpen
  \bibfield  {author} {\bibinfo {author} {\bibfnamefont {H.}~\bibnamefont
  {Reinken}}\ and\ \bibinfo {author} {\bibfnamefont {A.~M.}\ \bibnamefont
  {Menzel}},\ }\bibfield  {title} {\bibinfo {title} {Rheologically tuned modes
  of collective transport in active viscoelastic films},\ }\href@noop {}
  {\bibfield  {journal} {\bibinfo  {journal} {arXiv preprint arXiv:2502.06294}\
  } (\bibinfo {year} {2025}{\natexlab{b}})}\BibitemShut {NoStop}%
\bibitem [{\citenamefont {Pedley}\ and\ \citenamefont
  {Kessler}(1992)}]{pedley1992hydrodynamic}%
  \BibitemOpen
  \bibfield  {author} {\bibinfo {author} {\bibfnamefont {T.~J.}\ \bibnamefont
  {Pedley}}\ and\ \bibinfo {author} {\bibfnamefont {J.~O.}\ \bibnamefont
  {Kessler}},\ }\bibfield  {title} {\bibinfo {title} {Hydrodynamic phenomena in
  suspensions of swimming microorganisms},\ }\href@noop {} {\bibfield
  {journal} {\bibinfo  {journal} {Annu. Rev. Fluid Mech.}\ }\textbf {\bibinfo
  {volume} {24}},\ \bibinfo {pages} {313} (\bibinfo {year} {1992})}\BibitemShut
  {NoStop}%
\bibitem [{\citenamefont {Brotto}\ \emph {et~al.}(2013)\citenamefont {Brotto},
  \citenamefont {Caussin}, \citenamefont {Lauga},\ and\ \citenamefont
  {Bartolo}}]{brotto2013hydrodynamics}%
  \BibitemOpen
  \bibfield  {author} {\bibinfo {author} {\bibfnamefont {T.}~\bibnamefont
  {Brotto}}, \bibinfo {author} {\bibfnamefont {J.-B.}\ \bibnamefont {Caussin}},
  \bibinfo {author} {\bibfnamefont {E.}~\bibnamefont {Lauga}},\ and\ \bibinfo
  {author} {\bibfnamefont {D.}~\bibnamefont {Bartolo}},\ }\bibfield  {title}
  {\bibinfo {title} {Hydrodynamics of confined active fluids},\ }\href@noop {}
  {\bibfield  {journal} {\bibinfo  {journal} {Phys. Rev. Lett.}\ }\textbf
  {\bibinfo {volume} {110}},\ \bibinfo {pages} {038101} (\bibinfo {year}
  {2013})}\BibitemShut {NoStop}%
\bibitem [{\citenamefont {Maitra}\ \emph {et~al.}(2020)\citenamefont {Maitra},
  \citenamefont {Srivastava}, \citenamefont {Marchetti}, \citenamefont
  {Ramaswamy},\ and\ \citenamefont {Lenz}}]{maitra2020swimmer}%
  \BibitemOpen
  \bibfield  {author} {\bibinfo {author} {\bibfnamefont {A.}~\bibnamefont
  {Maitra}}, \bibinfo {author} {\bibfnamefont {P.}~\bibnamefont {Srivastava}},
  \bibinfo {author} {\bibfnamefont {M.~C.}\ \bibnamefont {Marchetti}}, \bibinfo
  {author} {\bibfnamefont {S.}~\bibnamefont {Ramaswamy}},\ and\ \bibinfo
  {author} {\bibfnamefont {M.}~\bibnamefont {Lenz}},\ }\bibfield  {title}
  {\bibinfo {title} {Swimmer suspensions on substrates: {A}nomalous stability
  and long-range order},\ }\href@noop {} {\bibfield  {journal} {\bibinfo
  {journal} {Phys. Rev. Lett.}\ }\textbf {\bibinfo {volume} {124}},\ \bibinfo
  {pages} {028002} (\bibinfo {year} {2020})}\BibitemShut {NoStop}%
\bibitem [{\citenamefont {Liu}\ \emph {et~al.}(2021)\citenamefont {Liu},
  \citenamefont {Shankar}, \citenamefont {Marchetti},\ and\ \citenamefont
  {Wu}}]{liu2021viscoelastic}%
  \BibitemOpen
  \bibfield  {author} {\bibinfo {author} {\bibfnamefont {S.}~\bibnamefont
  {Liu}}, \bibinfo {author} {\bibfnamefont {S.}~\bibnamefont {Shankar}},
  \bibinfo {author} {\bibfnamefont {M.~C.}\ \bibnamefont {Marchetti}},\ and\
  \bibinfo {author} {\bibfnamefont {Y.}~\bibnamefont {Wu}},\ }\bibfield
  {title} {\bibinfo {title} {Viscoelastic control of spatiotemporal order in
  bacterial active matter},\ }\href@noop {} {\bibfield  {journal} {\bibinfo
  {journal} {Nature}\ }\textbf {\bibinfo {volume} {590}},\ \bibinfo {pages}
  {80} (\bibinfo {year} {2021})}\BibitemShut {NoStop}%
\bibitem [{\citenamefont {Dadhichi}\ \emph {et~al.}(2018)\citenamefont
  {Dadhichi}, \citenamefont {Maitra},\ and\ \citenamefont
  {Ramaswamy}}]{dadhichi2018origins}%
  \BibitemOpen
  \bibfield  {author} {\bibinfo {author} {\bibfnamefont {L.~P.}\ \bibnamefont
  {Dadhichi}}, \bibinfo {author} {\bibfnamefont {A.}~\bibnamefont {Maitra}},\
  and\ \bibinfo {author} {\bibfnamefont {S.}~\bibnamefont {Ramaswamy}},\
  }\bibfield  {title} {\bibinfo {title} {Origins and diagnostics of the
  nonequilibrium character of active systems},\ }\href@noop {} {\bibfield
  {journal} {\bibinfo  {journal} {J. Stat. Mech. 123201\!\!}\ } (\bibinfo
  {year} {2018})}\BibitemShut {NoStop}%
\bibitem [{\citenamefont {Szab\'o}\ \emph {et~al.}(2006)\citenamefont
  {Szab\'o}, \citenamefont {Sz{\"o}ll{\"o}si}, \citenamefont {G{\"o}nci},
  \citenamefont {Jur{\'a}nyi}, \citenamefont {Selmeczi},\ and\ \citenamefont
  {Vicsek}}]{szabo2006phase}%
  \BibitemOpen
  \bibfield  {author} {\bibinfo {author} {\bibfnamefont {B.}~\bibnamefont
  {Szab\'o}}, \bibinfo {author} {\bibfnamefont {G.}~\bibnamefont
  {Sz{\"o}ll{\"o}si}}, \bibinfo {author} {\bibfnamefont {B.}~\bibnamefont
  {G{\"o}nci}}, \bibinfo {author} {\bibfnamefont {Z.}~\bibnamefont
  {Jur{\'a}nyi}}, \bibinfo {author} {\bibfnamefont {D.}~\bibnamefont
  {Selmeczi}},\ and\ \bibinfo {author} {\bibfnamefont {T.}~\bibnamefont
  {Vicsek}},\ }\bibfield  {title} {\bibinfo {title} {Phase transition in the
  collective migration of tissue cells: {E}xperiment and model},\ }\href@noop
  {} {\bibfield  {journal} {\bibinfo  {journal} {Phys. Rev. E}\ }\textbf
  {\bibinfo {volume} {74}},\ \bibinfo {pages} {061908} (\bibinfo {year}
  {2006})}\BibitemShut {NoStop}%
\bibitem [{\citenamefont {Lam}\ \emph {et~al.}(2015)\citenamefont {Lam},
  \citenamefont {Schindler},\ and\ \citenamefont {Dauchot}}]{lam2015self}%
  \BibitemOpen
  \bibfield  {author} {\bibinfo {author} {\bibfnamefont {K.-D. N.~T.}\
  \bibnamefont {Lam}}, \bibinfo {author} {\bibfnamefont {M.}~\bibnamefont
  {Schindler}},\ and\ \bibinfo {author} {\bibfnamefont {O.}~\bibnamefont
  {Dauchot}},\ }\bibfield  {title} {\bibinfo {title} {Self-propelled hard
  disks: {I}mplicit alignment and transition to collective motion},\
  }\href@noop {} {\bibfield  {journal} {\bibinfo  {journal} {New J. Phys.}\
  }\textbf {\bibinfo {volume} {17}},\ \bibinfo {pages} {113056} (\bibinfo
  {year} {2015})}\BibitemShut {NoStop}%
\bibitem [{\citenamefont {Baconnier}\ \emph {et~al.}(2025)\citenamefont
  {Baconnier}, \citenamefont {Dauchot}, \citenamefont {D{\'e}mery},
  \citenamefont {D{\"u}ring}, \citenamefont {Henkes}, \citenamefont {Huepe},\
  and\ \citenamefont {Shee}}]{baconnier2025self}%
  \BibitemOpen
  \bibfield  {author} {\bibinfo {author} {\bibfnamefont {P.}~\bibnamefont
  {Baconnier}}, \bibinfo {author} {\bibfnamefont {O.}~\bibnamefont {Dauchot}},
  \bibinfo {author} {\bibfnamefont {V.}~\bibnamefont {D{\'e}mery}}, \bibinfo
  {author} {\bibfnamefont {G.}~\bibnamefont {D{\"u}ring}}, \bibinfo {author}
  {\bibfnamefont {S.}~\bibnamefont {Henkes}}, \bibinfo {author} {\bibfnamefont
  {C.}~\bibnamefont {Huepe}},\ and\ \bibinfo {author} {\bibfnamefont
  {A.}~\bibnamefont {Shee}},\ }\bibfield  {title} {\bibinfo {title}
  {Self-aligning polar active matter},\ }\href@noop {} {\bibfield  {journal}
  {\bibinfo  {journal} {Rev. Mod. Phys.}\ }\textbf {\bibinfo {volume} {97}},\
  \bibinfo {pages} {015007} (\bibinfo {year} {2025})}\BibitemShut {NoStop}%
\bibitem [{\citenamefont {Risken}(1984)}]{risken1984fokker}%
  \BibitemOpen
  \bibfield  {author} {\bibinfo {author} {\bibfnamefont {H.}~\bibnamefont
  {Risken}},\ }\href@noop {} {\emph {\bibinfo {title} {The Fokker-Planck
  Equation}}}\ (\bibinfo  {publisher} {Springer, Berlin Heidelberg},\ \bibinfo
  {year} {1984})\BibitemShut {NoStop}%
\bibitem [{\citenamefont {Doi}\ and\ \citenamefont
  {Edwards}(1988)}]{doi1988theory}%
  \BibitemOpen
  \bibfield  {author} {\bibinfo {author} {\bibfnamefont {M.}~\bibnamefont
  {Doi}}\ and\ \bibinfo {author} {\bibfnamefont {S.~F.}\ \bibnamefont
  {Edwards}},\ }\href@noop {} {\emph {\bibinfo {title} {The Theory of Polymer
  Dynamics}}}\ (\bibinfo  {publisher} {Oxford University Press, Oxford},\
  \bibinfo {year} {1988})\BibitemShut {NoStop}%
\bibitem [{\citenamefont {Rin{\"a}cker}\ and\ \citenamefont
  {Hess}(1998)}]{rienaecker1998orientational}%
  \BibitemOpen
  \bibfield  {author} {\bibinfo {author} {\bibfnamefont {G.}~\bibnamefont
  {Rin{\"a}cker}}\ and\ \bibinfo {author} {\bibfnamefont {S.}~\bibnamefont
  {Hess}},\ }\bibfield  {title} {\bibinfo {title} {Orientational dynamics of
  nematic liquid crystals under shear flow},\ }\href@noop {} {\bibfield
  {journal} {\bibinfo  {journal} {Physica A}\ }\textbf {\bibinfo {volume}
  {267}},\ \bibinfo {pages} {294} (\bibinfo {year} {1998})}\BibitemShut
  {NoStop}%
\bibitem [{\citenamefont {Kr{\"o}ger}\ \emph {et~al.}(2008)\citenamefont
  {Kr{\"o}ger}, \citenamefont {Ammar},\ and\ \citenamefont
  {Chinesta}}]{kroeger2008consistent}%
  \BibitemOpen
  \bibfield  {author} {\bibinfo {author} {\bibfnamefont {M.}~\bibnamefont
  {Kr{\"o}ger}}, \bibinfo {author} {\bibfnamefont {A.}~\bibnamefont {Ammar}},\
  and\ \bibinfo {author} {\bibfnamefont {F.}~\bibnamefont {Chinesta}},\
  }\bibfield  {title} {\bibinfo {title} {Consistent closure schemes for
  statistical models of anisotropic fluids},\ }\href@noop {} {\bibfield
  {journal} {\bibinfo  {journal} {J. Non-{N}ewton. Fluid Mech.}\ }\textbf
  {\bibinfo {volume} {149}},\ \bibinfo {pages} {50} (\bibinfo {year}
  {2008})}\BibitemShut {NoStop}%
\bibitem [{\citenamefont {Hand}(1962)}]{hand1962theory}%
  \BibitemOpen
  \bibfield  {author} {\bibinfo {author} {\bibfnamefont {G.~L.}\ \bibnamefont
  {Hand}},\ }\bibfield  {title} {\bibinfo {title} {A theory of anisotropic
  fluids},\ }\href@noop {} {\bibfield  {journal} {\bibinfo  {journal} {J. Fluid
  Mech.}\ }\textbf {\bibinfo {volume} {13}},\ \bibinfo {pages} {33} (\bibinfo
  {year} {1962})}\BibitemShut {NoStop}%
\bibitem [{\citenamefont {Martin}\ \emph {et~al.}(1972)\citenamefont {Martin},
  \citenamefont {Parodi},\ and\ \citenamefont {Pershan}}]{martin1972unified}%
  \BibitemOpen
  \bibfield  {author} {\bibinfo {author} {\bibfnamefont {P.~C.}\ \bibnamefont
  {Martin}}, \bibinfo {author} {\bibfnamefont {O.}~\bibnamefont {Parodi}},\
  and\ \bibinfo {author} {\bibfnamefont {P.~S.}\ \bibnamefont {Pershan}},\
  }\bibfield  {title} {\bibinfo {title} {Unified hydrodynamic theory for
  crystals, liquid crystals, and normal fluids},\ }\href@noop {} {\bibfield
  {journal} {\bibinfo  {journal} {Phys. Rev. A}\ }\textbf {\bibinfo {volume}
  {6}},\ \bibinfo {pages} {2401} (\bibinfo {year} {1972})}\BibitemShut
  {NoStop}%
\bibitem [{\citenamefont {Temmen}\ \emph {et~al.}(2000)\citenamefont {Temmen},
  \citenamefont {Pleiner}, \citenamefont {Liu},\ and\ \citenamefont
  {Brand}}]{temmen2000convective}%
  \BibitemOpen
  \bibfield  {author} {\bibinfo {author} {\bibfnamefont {H.}~\bibnamefont
  {Temmen}}, \bibinfo {author} {\bibfnamefont {H.}~\bibnamefont {Pleiner}},
  \bibinfo {author} {\bibfnamefont {M.}~\bibnamefont {Liu}},\ and\ \bibinfo
  {author} {\bibfnamefont {H.~R.}\ \bibnamefont {Brand}},\ }\bibfield  {title}
  {\bibinfo {title} {Convective nonlinearity in non-{N}ewtonian fluids},\
  }\href@noop {} {\bibfield  {journal} {\bibinfo  {journal} {Phys. Rev. Lett.}\
  }\textbf {\bibinfo {volume} {84}},\ \bibinfo {pages} {3228} (\bibinfo {year}
  {2000})}\BibitemShut {NoStop}%
\bibitem [{\citenamefont {Puljiz}\ and\ \citenamefont
  {Menzel}(2019)}]{puljiz2019memory}%
  \BibitemOpen
  \bibfield  {author} {\bibinfo {author} {\bibfnamefont {M.}~\bibnamefont
  {Puljiz}}\ and\ \bibinfo {author} {\bibfnamefont {A.~M.}\ \bibnamefont
  {Menzel}},\ }\bibfield  {title} {\bibinfo {title} {Memory-based mediated
  interactions between rigid particulate inclusions in viscoelastic
  environments},\ }\href@noop {} {\bibfield  {journal} {\bibinfo  {journal}
  {Phys. Rev. E}\ }\textbf {\bibinfo {volume} {99}},\ \bibinfo {pages} {012601}
  (\bibinfo {year} {2019})}\BibitemShut {NoStop}%
\bibitem [{\citenamefont {Landau}\ and\ \citenamefont
  {Lifshitz}(1987)}]{landau1987fluid}%
  \BibitemOpen
  \bibfield  {author} {\bibinfo {author} {\bibfnamefont {L.~D.}\ \bibnamefont
  {Landau}}\ and\ \bibinfo {author} {\bibfnamefont {E.~M.}\ \bibnamefont
  {Lifshitz}},\ }\href@noop {} {\emph {\bibinfo {title} {Fluid Mechanics}}}\
  (\bibinfo  {publisher} {Elsevier, Amsterdam},\ \bibinfo {year}
  {1987})\BibitemShut {NoStop}%
\bibitem [{\citenamefont {Landau}\ and\ \citenamefont
  {Lifshitz}(1986)}]{landau1986theory}%
  \BibitemOpen
  \bibfield  {author} {\bibinfo {author} {\bibfnamefont {L.~D.}\ \bibnamefont
  {Landau}}\ and\ \bibinfo {author} {\bibfnamefont {E.~M.}\ \bibnamefont
  {Lifshitz}},\ }\href@noop {} {\emph {\bibinfo {title} {Theory of
  Elasticity}}}\ (\bibinfo  {publisher} {Butterworth-Heinemann, Oxford},\
  \bibinfo {year} {1986})\BibitemShut {NoStop}%
\bibitem [{\citenamefont {Chaikin}\ \emph {et~al.}(1995)\citenamefont
  {Chaikin}, \citenamefont {Lubensky},\ and\ \citenamefont
  {Witten}}]{chaikin1995principles}%
  \BibitemOpen
  \bibfield  {author} {\bibinfo {author} {\bibfnamefont {P.~M.}\ \bibnamefont
  {Chaikin}}, \bibinfo {author} {\bibfnamefont {T.~C.}\ \bibnamefont
  {Lubensky}},\ and\ \bibinfo {author} {\bibfnamefont {T.~A.}\ \bibnamefont
  {Witten}},\ }\href@noop {} {\emph {\bibinfo {title} {Principles of Condensed
  Matter Physics}}},\ Vol.~\bibinfo {volume} {1}\ (\bibinfo  {publisher}
  {Cambridge University Press},\ \bibinfo {year} {1995})\BibitemShut {NoStop}%
\bibitem [{\citenamefont {Pleiner}\ \emph {et~al.}(2004)\citenamefont
  {Pleiner}, \citenamefont {Liu},\ and\ \citenamefont
  {Brand}}]{pleiner2004nonlinear}%
  \BibitemOpen
  \bibfield  {author} {\bibinfo {author} {\bibfnamefont {H.}~\bibnamefont
  {Pleiner}}, \bibinfo {author} {\bibfnamefont {M.}~\bibnamefont {Liu}},\ and\
  \bibinfo {author} {\bibfnamefont {H.~R.}\ \bibnamefont {Brand}},\ }\bibfield
  {title} {\bibinfo {title} {Nonlinear fluid dynamics description of
  non-{N}ewtonian fluids},\ }\href@noop {} {\bibfield  {journal} {\bibinfo
  {journal} {Rheol. Acta}\ }\textbf {\bibinfo {volume} {43}},\ \bibinfo {pages}
  {502} (\bibinfo {year} {2004})}\BibitemShut {NoStop}%
\bibitem [{\citenamefont {Xu}\ \emph {et~al.}(2023)\citenamefont {Xu},
  \citenamefont {Huang}, \citenamefont {Zhang},\ and\ \citenamefont
  {Wu}}]{xu2023autonomous}%
  \BibitemOpen
  \bibfield  {author} {\bibinfo {author} {\bibfnamefont {H.}~\bibnamefont
  {Xu}}, \bibinfo {author} {\bibfnamefont {Y.}~\bibnamefont {Huang}}, \bibinfo
  {author} {\bibfnamefont {R.}~\bibnamefont {Zhang}},\ and\ \bibinfo {author}
  {\bibfnamefont {Y.}~\bibnamefont {Wu}},\ }\bibfield  {title} {\bibinfo
  {title} {Autonomous waves and global motion modes in living active solids},\
  }\href@noop {} {\bibfield  {journal} {\bibinfo  {journal} {Nature Phys.}\
  }\textbf {\bibinfo {volume} {19}},\ \bibinfo {pages} {46} (\bibinfo {year}
  {2023})}\BibitemShut {NoStop}%
\bibitem [{\citenamefont {Simha}\ and\ \citenamefont
  {Ramaswamy}(2002)}]{simha2002hydrodynamic}%
  \BibitemOpen
  \bibfield  {author} {\bibinfo {author} {\bibfnamefont {R.~A.}\ \bibnamefont
  {Simha}}\ and\ \bibinfo {author} {\bibfnamefont {S.}~\bibnamefont
  {Ramaswamy}},\ }\bibfield  {title} {\bibinfo {title} {Hydrodynamic
  fluctuations and instabilities in ordered suspensions of self-propelled
  particles},\ }\href@noop {} {\bibfield  {journal} {\bibinfo  {journal} {Phys.
  Rev. Lett.}\ }\textbf {\bibinfo {volume} {89}},\ \bibinfo {pages} {058101}
  (\bibinfo {year} {2002})}\BibitemShut {NoStop}%
\bibitem [{\citenamefont {Saintillan}\ and\ \citenamefont
  {Shelley}(2008)}]{saintillan2008instabilities}%
  \BibitemOpen
  \bibfield  {author} {\bibinfo {author} {\bibfnamefont {D.}~\bibnamefont
  {Saintillan}}\ and\ \bibinfo {author} {\bibfnamefont {M.~J.}\ \bibnamefont
  {Shelley}},\ }\bibfield  {title} {\bibinfo {title} {Instabilities and pattern
  formation in active particle suspensions: {K}inetic theory and continuum
  simulations},\ }\href@noop {} {\bibfield  {journal} {\bibinfo  {journal}
  {Phys. Rev. Lett.}\ }\textbf {\bibinfo {volume} {100}},\ \bibinfo {pages}
  {178103} (\bibinfo {year} {2008})}\BibitemShut {NoStop}%
\bibitem [{\citenamefont {Richter}\ \emph {et~al.}(2021)\citenamefont
  {Richter}, \citenamefont {Deters},\ and\ \citenamefont
  {Menzel}}]{richter2021rotating}%
  \BibitemOpen
  \bibfield  {author} {\bibinfo {author} {\bibfnamefont {S.~K.}\ \bibnamefont
  {Richter}}, \bibinfo {author} {\bibfnamefont {C.~D.}\ \bibnamefont
  {Deters}},\ and\ \bibinfo {author} {\bibfnamefont {A.~M.}\ \bibnamefont
  {Menzel}},\ }\bibfield  {title} {\bibinfo {title} {Rotating spherical
  particle in a continuous viscoelastic medium—a microrheological example
  situation},\ }\href@noop {} {\bibfield  {journal} {\bibinfo  {journal}
  {Europhys. Lett.}\ }\textbf {\bibinfo {volume} {134}},\ \bibinfo {pages}
  {68002} (\bibinfo {year} {2021})}\BibitemShut {NoStop}%
\bibitem [{\citenamefont {de~Gennes}(1979)}]{de1979scaling}%
  \BibitemOpen
  \bibfield  {author} {\bibinfo {author} {\bibfnamefont {P.-G.}\ \bibnamefont
  {de~Gennes}},\ }\href@noop {} {\emph {\bibinfo {title} {Scaling Concepts in
  Polymer Physics}}}\ (\bibinfo  {publisher} {Cornell University Press,
  Ithaca},\ \bibinfo {year} {1979})\BibitemShut {NoStop}%
\bibitem [{\citenamefont {Menzel}(2025)}]{menzel2025linear}%
  \BibitemOpen
  \bibfield  {author} {\bibinfo {author} {\bibfnamefont {A.~M.}\ \bibnamefont
  {Menzel}},\ }\bibfield  {title} {\bibinfo {title} {Linear theory of
  viscoelasticity in a generalized hydrodynamic framework},\ }\href@noop {}
  {\bibfield  {journal} {\bibinfo  {journal} {arXiv preprint arXiv:2505.10032}\
  } (\bibinfo {year} {2025})}\BibitemShut {NoStop}%
\bibitem [{\citenamefont {Nijjer}\ \emph {et~al.}(2023)\citenamefont {Nijjer},
  \citenamefont {Cohen},\ and\ \citenamefont {Yan}}]{nijjer2023bacteria}%
  \BibitemOpen
  \bibfield  {author} {\bibinfo {author} {\bibfnamefont {J.}~\bibnamefont
  {Nijjer}}, \bibinfo {author} {\bibfnamefont {T.}~\bibnamefont {Cohen}},\ and\
  \bibinfo {author} {\bibfnamefont {J.}~\bibnamefont {Yan}},\ }\bibfield
  {title} {\bibinfo {title} {Bacteria surfing the elastic wave},\ }\href@noop
  {} {\bibfield  {journal} {\bibinfo  {journal} {Nature Phys.}\ }\textbf
  {\bibinfo {volume} {19}},\ \bibinfo {pages} {6} (\bibinfo {year}
  {2023})}\BibitemShut {NoStop}%
\bibitem [{\citenamefont {Zhang}\ \emph {et~al.}(2021)\citenamefont {Zhang},
  \citenamefont {Alert}, \citenamefont {Yan}, \citenamefont {Wingreen},\ and\
  \citenamefont {Granick}}]{zhang2021active}%
  \BibitemOpen
  \bibfield  {author} {\bibinfo {author} {\bibfnamefont {J.}~\bibnamefont
  {Zhang}}, \bibinfo {author} {\bibfnamefont {R.}~\bibnamefont {Alert}},
  \bibinfo {author} {\bibfnamefont {J.}~\bibnamefont {Yan}}, \bibinfo {author}
  {\bibfnamefont {N.~S.}\ \bibnamefont {Wingreen}},\ and\ \bibinfo {author}
  {\bibfnamefont {S.}~\bibnamefont {Granick}},\ }\bibfield  {title} {\bibinfo
  {title} {Active phase separation by turning towards regions of higher
  density},\ }\href@noop {} {\bibfield  {journal} {\bibinfo  {journal} {Nat.
  Phys.}\ }\textbf {\bibinfo {volume} {17}},\ \bibinfo {pages} {961} (\bibinfo
  {year} {2021})}\BibitemShut {NoStop}%
\bibitem [{\citenamefont {Das}\ \emph {et~al.}(2024)\citenamefont {Das},
  \citenamefont {Ciarchi}, \citenamefont {Zhou}, \citenamefont {Yan},
  \citenamefont {Zhang},\ and\ \citenamefont {Alert}}]{das2024flocking}%
  \BibitemOpen
  \bibfield  {author} {\bibinfo {author} {\bibfnamefont {S.}~\bibnamefont
  {Das}}, \bibinfo {author} {\bibfnamefont {M.}~\bibnamefont {Ciarchi}},
  \bibinfo {author} {\bibfnamefont {Z.}~\bibnamefont {Zhou}}, \bibinfo {author}
  {\bibfnamefont {J.}~\bibnamefont {Yan}}, \bibinfo {author} {\bibfnamefont
  {J.}~\bibnamefont {Zhang}},\ and\ \bibinfo {author} {\bibfnamefont
  {R.}~\bibnamefont {Alert}},\ }\bibfield  {title} {\bibinfo {title} {Flocking
  by turning away},\ }\href@noop {} {\bibfield  {journal} {\bibinfo  {journal}
  {Phys. Rev. X}\ }\textbf {\bibinfo {volume} {14}},\ \bibinfo {pages} {031008}
  (\bibinfo {year} {2024})}\BibitemShut {NoStop}%
\bibitem [{\citenamefont {Casiulis}\ \emph {et~al.}(2024)\citenamefont
  {Casiulis}, \citenamefont {Arbel}, \citenamefont {Lahini}, \citenamefont
  {Martiniani}, \citenamefont {Oppenheimer},\ and\ \citenamefont
  {Zion}}]{casiulis2024geometric}%
  \BibitemOpen
  \bibfield  {author} {\bibinfo {author} {\bibfnamefont {M.}~\bibnamefont
  {Casiulis}}, \bibinfo {author} {\bibfnamefont {E.}~\bibnamefont {Arbel}},
  \bibinfo {author} {\bibfnamefont {Y.}~\bibnamefont {Lahini}}, \bibinfo
  {author} {\bibfnamefont {S.}~\bibnamefont {Martiniani}}, \bibinfo {author}
  {\bibfnamefont {N.}~\bibnamefont {Oppenheimer}},\ and\ \bibinfo {author}
  {\bibfnamefont {M.~Y.~B.}\ \bibnamefont {Zion}},\ }\bibfield  {title}
  {\bibinfo {title} {A geometric condition for robot-swarm cohesion and
  cluster-flock transition},\ }\href@noop {} {\bibfield  {journal} {\bibinfo
  {journal} {arXiv preprint arXiv:2409.04618}\ } (\bibinfo {year}
  {2024})}\BibitemShut {NoStop}%
\bibitem [{\citenamefont {Arbel}\ \emph {et~al.}(2024)\citenamefont {Arbel},
  \citenamefont {Oppenheimer}, \citenamefont {Lahini},\ and\ \citenamefont
  {Zion}}]{arbel2024mechanical}%
  \BibitemOpen
  \bibfield  {author} {\bibinfo {author} {\bibfnamefont {E.}~\bibnamefont
  {Arbel}}, \bibinfo {author} {\bibfnamefont {N.}~\bibnamefont {Oppenheimer}},
  \bibinfo {author} {\bibfnamefont {Y.}~\bibnamefont {Lahini}},\ and\ \bibinfo
  {author} {\bibfnamefont {M.~Y.~B.}\ \bibnamefont {Zion}},\ }\bibfield
  {title} {\bibinfo {title} {A mechanical origin of cooperative transport},\
  }\href@noop {} {\bibfield  {journal} {\bibinfo  {journal} {arXiv preprint
  arXiv:2402.05659}\ } (\bibinfo {year} {2024})}\BibitemShut {NoStop}%
\bibitem [{\citenamefont {Liebchen}\ and\ \citenamefont
  {Levis}(2022)}]{liebchen2022chiral}%
  \BibitemOpen
  \bibfield  {author} {\bibinfo {author} {\bibfnamefont {B.}~\bibnamefont
  {Liebchen}}\ and\ \bibinfo {author} {\bibfnamefont {D.}~\bibnamefont
  {Levis}},\ }\bibfield  {title} {\bibinfo {title} {Chiral active matter},\
  }\href@noop {} {\bibfield  {journal} {\bibinfo  {journal} {Europhys. Lett.}\
  }\textbf {\bibinfo {volume} {139}},\ \bibinfo {pages} {67001} (\bibinfo
  {year} {2022})}\BibitemShut {NoStop}%
\bibitem [{\citenamefont {Lei}\ \emph {et~al.}(2019)\citenamefont {Lei},
  \citenamefont {Ciamarra},\ and\ \citenamefont {Ni}}]{lei2019nonequilibrium}%
  \BibitemOpen
  \bibfield  {author} {\bibinfo {author} {\bibfnamefont {Q.-L.}\ \bibnamefont
  {Lei}}, \bibinfo {author} {\bibfnamefont {M.~P.}\ \bibnamefont {Ciamarra}},\
  and\ \bibinfo {author} {\bibfnamefont {R.}~\bibnamefont {Ni}},\ }\bibfield
  {title} {\bibinfo {title} {Nonequilibrium strongly hyperuniform fluids of
  circle active particles with large local density fluctuations},\ }\href@noop
  {} {\bibfield  {journal} {\bibinfo  {journal} {Sci. Adv.}\ }\textbf {\bibinfo
  {volume} {5}},\ \bibinfo {pages} {eaau7423} (\bibinfo {year}
  {2019})}\BibitemShut {NoStop}%
\bibitem [{\citenamefont {Menzel}\ \emph {et~al.}(2014)\citenamefont {Menzel},
  \citenamefont {Ohta},\ and\ \citenamefont {L{\"o}wen}}]{menzel2014active}%
  \BibitemOpen
  \bibfield  {author} {\bibinfo {author} {\bibfnamefont {A.~M.}\ \bibnamefont
  {Menzel}}, \bibinfo {author} {\bibfnamefont {T.}~\bibnamefont {Ohta}},\ and\
  \bibinfo {author} {\bibfnamefont {H.}~\bibnamefont {L{\"o}wen}},\ }\bibfield
  {title} {\bibinfo {title} {Active crystals and their stability},\ }\href@noop
  {} {\bibfield  {journal} {\bibinfo  {journal} {Phys. Rev. E}\ }\textbf
  {\bibinfo {volume} {89}},\ \bibinfo {pages} {022301} (\bibinfo {year}
  {2014})}\BibitemShut {NoStop}%
\bibitem [{\citenamefont {Hoell}\ \emph {et~al.}(2019)\citenamefont {Hoell},
  \citenamefont {L{\"o}wen},\ and\ \citenamefont {Menzel}}]{hoell2019multi}%
  \BibitemOpen
  \bibfield  {author} {\bibinfo {author} {\bibfnamefont {C.}~\bibnamefont
  {Hoell}}, \bibinfo {author} {\bibfnamefont {H.}~\bibnamefont {L{\"o}wen}},\
  and\ \bibinfo {author} {\bibfnamefont {A.~M.}\ \bibnamefont {Menzel}},\
  }\bibfield  {title} {\bibinfo {title} {Multi-species dynamical density
  functional theory for microswimmers: Derivation, orientational ordering,
  trapping potentials, and shear cells},\ }\href@noop {} {\bibfield  {journal}
  {\bibinfo  {journal} {J. Chem. Phys.}\ }\textbf {\bibinfo {volume} {151}}
  (\bibinfo {year} {2019})}\BibitemShut {NoStop}%
\bibitem [{\citenamefont {Bertin}\ \emph {et~al.}(2006)\citenamefont {Bertin},
  \citenamefont {Droz},\ and\ \citenamefont
  {Gr{\'e}goire}}]{bertin2006boltzmann}%
  \BibitemOpen
  \bibfield  {author} {\bibinfo {author} {\bibfnamefont {E.}~\bibnamefont
  {Bertin}}, \bibinfo {author} {\bibfnamefont {M.}~\bibnamefont {Droz}},\ and\
  \bibinfo {author} {\bibfnamefont {G.}~\bibnamefont {Gr{\'e}goire}},\
  }\bibfield  {title} {\bibinfo {title} {Boltzmann and hydrodynamic description
  for self-propelled particles},\ }\href@noop {} {\bibfield  {journal}
  {\bibinfo  {journal} {Phys. Rev. E}\ }\textbf {\bibinfo {volume} {74}},\
  \bibinfo {pages} {022101} (\bibinfo {year} {2006})}\BibitemShut {NoStop}%
\bibitem [{\citenamefont {Bertin}\ \emph {et~al.}(2009)\citenamefont {Bertin},
  \citenamefont {Droz},\ and\ \citenamefont
  {Gr{\'e}goire}}]{bertin2009hydrodynamic}%
  \BibitemOpen
  \bibfield  {author} {\bibinfo {author} {\bibfnamefont {E.}~\bibnamefont
  {Bertin}}, \bibinfo {author} {\bibfnamefont {M.}~\bibnamefont {Droz}},\ and\
  \bibinfo {author} {\bibfnamefont {G.}~\bibnamefont {Gr{\'e}goire}},\
  }\bibfield  {title} {\bibinfo {title} {Hydrodynamic equations for
  self-propelled particles: microscopic derivation and stability analysis},\
  }\href@noop {} {\bibfield  {journal} {\bibinfo  {journal} {J. Phys. A}\
  }\textbf {\bibinfo {volume} {42}},\ \bibinfo {pages} {445001} (\bibinfo
  {year} {2009})}\BibitemShut {NoStop}%
\bibitem [{\citenamefont {Dean}(1996)}]{dean1996langevin}%
  \BibitemOpen
  \bibfield  {author} {\bibinfo {author} {\bibfnamefont {D.~S.}\ \bibnamefont
  {Dean}},\ }\bibfield  {title} {\bibinfo {title} {Langevin equation for the
  density of a system of interacting {L}angevin processes},\ }\href@noop {}
  {\bibfield  {journal} {\bibinfo  {journal} {J. Phys. A}\ }\textbf {\bibinfo
  {volume} {29}},\ \bibinfo {pages} {L613} (\bibinfo {year}
  {1996})}\BibitemShut {NoStop}%
\bibitem [{\citenamefont {Reinken}(2025)}]{code2025unified}%
  \BibitemOpen
  \bibfield  {author} {\bibinfo {author} {\bibfnamefont {H.}~\bibnamefont
  {Reinken}},\ }\href@noop {} {\bibinfo {title} {Github repository}},\ \bibinfo
  {howpublished} {https://github.com/
  henningreinken/Unified-description-of-viscous-viscoelastic-or-elastic-thin-active-films-on-substrates}
  (\bibinfo {year} {2025})\BibitemShut {NoStop}%
\bibitem [{\citenamefont {Strobl}(1997)}]{strobl1997physics}%
  \BibitemOpen
  \bibfield  {author} {\bibinfo {author} {\bibfnamefont {G.~R.}\ \bibnamefont
  {Strobl}},\ }\href@noop {} {\emph {\bibinfo {title} {The Physics of
  Polymers}}}\ (\bibinfo  {publisher} {Springer, Berlin Heidelberg},\ \bibinfo
  {year} {1997})\BibitemShut {NoStop}%
\bibitem [{\citenamefont {Canuto}\ \emph {et~al.}(2007)\citenamefont {Canuto},
  \citenamefont {Hussaini}, \citenamefont {Quarteroni},\ and\ \citenamefont
  {Zang}}]{canuto2007spectral}%
  \BibitemOpen
  \bibfield  {author} {\bibinfo {author} {\bibfnamefont {C.}~\bibnamefont
  {Canuto}}, \bibinfo {author} {\bibfnamefont {M.~Y.}\ \bibnamefont
  {Hussaini}}, \bibinfo {author} {\bibfnamefont {A.}~\bibnamefont
  {Quarteroni}},\ and\ \bibinfo {author} {\bibfnamefont {T.~A.}\ \bibnamefont
  {Zang}},\ }\href@noop {} {\emph {\bibinfo {title} {Spectral Methods:
  {E}volution to Complex Geometries and Applications to Fluid Dynamics}}}\
  (\bibinfo  {publisher} {Springer, Berlin Heidelberg},\ \bibinfo {year}
  {2007})\BibitemShut {NoStop}%
\bibitem [{\citenamefont {Durran}(2010)}]{durran2010numerical}%
  \BibitemOpen
  \bibfield  {author} {\bibinfo {author} {\bibfnamefont {D.~R.}\ \bibnamefont
  {Durran}},\ }\href@noop {} {\emph {\bibinfo {title} {Numerical Methods for
  Fluid Dynamics}}}\ (\bibinfo  {publisher} {Springer, New York},\ \bibinfo
  {year} {2010})\BibitemShut {NoStop}%
\end{thebibliography}

%

\end{document}